\documentclass[11pt]{article}

\usepackage[ruled,vlined]{algorithm2e}
\usepackage{fullpage}

\usepackage{tikz}
\usepackage{amsmath}
\usepackage{xspace}
\usepackage{color}
\usepackage{colortbl}
\usepackage{multirow}
\usepackage{hhline}
\usepackage{subfigure}
\usepackage{comment}
\usepackage[ruled,vlined]{algorithm2e}
\usepackage[font={small,bf}]{caption}
\usepackage[normalem]{ulem}
\PassOptionsToPackage{hyphens}{url}
\usepackage{hyperref}
\usepackage{arydshln}
\usepackage{wrapfig}
\usepackage[small,compact]{titlesec}

\usepackage{fancyhdr}
\newcommand{\mysection}[1]{\section{\uppercase{#1}}}
\newcommand{\mycomment}[1]{}

\newcommand{\newtext}[1]{#1}
\renewcommand{\vec}[1]{\overrightarrow{#1}}

\newcommand{\ie}{\textit{i.e.,}}
\newcommand{\eg}{\textit{e.g.,}}
\newcommand{\etal}{\textit{et al.}}

\newcommand{\naive}{na\"{i}ve\xspace}

\newcommand{\circled}[1]{\tikz[baseline=(char.base)]{
            \node[shape=circle,draw,inner sep=0.5pt] (char) {#1};}}

\newcommand{\bfcircled}[1]{\circled{\textbf{#1}}}

\newcommand{\sectref}[1]{Section~\ref{#1}}
\newcommand{\appref}[1]{Appendix~\ref{#1}}
\newcommand{\figref}[1]{Figure~\ref{#1}}

\newcommand{\anonsect}[1]{\par\noindent\textbf{\textsf{#1}}}

\newcommand{\mylineref}[1]{\ref{#1}}
\newcommand{\MPCALGOLINEPREFIX}{M}
\newcommand{\mpclref}[1]{\MPCALGOLINEPREFIX\ref{#1}}
\newcommand{\GCALGOLINEPREFIX}{GC}
\newcommand{\gclref}[1]{\GCALGOLINEPREFIX\ref{#1}}

\newcounter{myctr}
\newenvironment{mylist}{\begin{list}{\textbf{\bfcircled{\arabic{myctr}}}}
{\usecounter{myctr}
\setlength{\topsep}{1mm}\setlength{\itemsep}{0.5mm}
\setlength{\parsep}{0.5mm}
\setlength{\listparindent}{\parindent} 
\setlength{\itemindent}{0mm}\setlength{\partopsep}{0mm}
\setlength{\labelwidth}{-2mm}
\setlength{\leftmargin}{0mm}}}{\end{list}}

\newenvironment{mybullet}{\begin{list}{$\bullet$}
{\setlength{\topsep}{1mm}\setlength{\itemsep}{0.5mm}
\setlength{\parsep}{0.5mm}
\setlength{\listparindent}{\parindent} 
\setlength{\itemindent}{0mm}\setlength{\partopsep}{0mm}
\setlength{\labelwidth}{-2mm}
\setlength{\leftmargin}{0mm}}}{\end{list}}

\definecolor{LightGray}{gray}{0.97}

\newcommand{\negspace}[1]{\indent\vspace{#1}}
\newcommand{\negspaceqtr}{\negspace{-0.25cm}}
\newcommand{\negspacehalf}{\negspace{-0.5cm}}
\newcommand{\negspacetqtr}{\negspace{-0.75cm}}

\newcommand{\pd}{Privadome\xspace}
\newcommand{\pdmpc}{\textsc{\uppercase{Pd-Mpc}}\xspace}
\newcommand{\pdros}{\textsc{\uppercase{Pd-Ros}}\xspace}
\newcommand{\tplusd}{{t\texttt{+}\delta}}
\newcommand{\funcname}[1]{\textsc{#1}}
\newcommand{\rostopic}[1]{\textsf{\textit{#1}}}
\newcommand{\latitude}{\Gamma}
\newcommand{\longitude}{\lambda}
\newcommand{\goal}[1]{\textbf{\footnotesize #1}}
\newcommand{\tabdataref}[1]{#1}

\newcommand{\lan}{\textsc{\small lan}}
\newcommand{\wan}{\textsc{\small wan}}

\makeatletter
\def\thickhline{%
  \noalign{\ifnum0=`}\fi\hrule \@height \thickarrayrulewidth \futurelet
   \reserved@a\@xthickhline}
\def\@xthickhline{\ifx\reserved@a\thickhline
               \vskip\doublerulesep
               \vskip-\thickarrayrulewidth
             \fi
      \ifnum0=`{\fi}}
\makeatother

\newlength{\thickarrayrulewidth}
\usepackage{amsmath}
\usepackage{amsfonts}

\renewcommand{\arctan}{\texttt{arctan}}
\renewcommand{\arccos}{\texttt{arccos}}

\title{\bf \pd: Protecting Citizen Privacy\\from Delivery Drones}
\author{Gokulnath Pillai, Eikansh Gupta, Ajith Suresh, Vinod Ganapathy and Arpita Patra\\
Indian Institute of Science, Bangalore}
\date{May 2022}

\begin{document}

\hypersetup{breaklinks=true,
            hidelinks=true,
            linkcolor=black,
            citecolor=black}
\renewcommand{\UrlFont}{\sffamily\footnotesize}
\maketitle

\begin{abstract}
    
As e-commerce companies begin to consider using delivery drones for customer fulfillment,  there are growing concerns around citizen privacy. Drones are equipped with cameras, and the video feed from these cameras is often required as part of routine navigation, be it for semi-autonomous or fully-autonomous drones. Footage of ground-based citizens may be captured in this video feed, thereby leading to privacy concerns.
    
This paper presents \pd, a system that implements the vision of a virtual privacy dome centered around the citizen. \pd\ is designed to be integrated with city-scale regulatory authorities that oversee delivery drone operations and realizes this vision through two components, \pdmpc\ and \pdros. \pdmpc\ allows citizens equipped with a mobile device to identify drones that have captured their footage. It uses secure two-party computation to achieve this goal without compromising the privacy of the citizen's location. 
\pdros\ allows the citizen to communicate with such drones and obtain an audit trail showing how the drone uses their footage and determine if privacy-preserving steps are taken to sanitize the footage. An experimental evaluation of \pd\ using our prototype implementations of \pdmpc\ and \pdros\ shows that the system scales to near-term city-scale delivery drone deployments (hundreds of drones). We show that with \pdmpc\ the mobile data usage on the citizen's mobile device is comparable to that of routine activities on the device, such as streaming videos. We also show that the workflow of \pdros\ consumes a modest amount of additional CPU resources and power on our experimental platform.
\end{abstract}

\mysection{Introduction}
\label{sec:intro}


This paper concerns the problem of citizen privacy in the era of delivery drones. Prior studies~\cite{spiders:chi2017,bystanders:chi2017,perceptions:popets2016,nassi:oakland21} have shown that citizens perceive drones as a threat to their privacy, and rightly so. Drones are equipped with cameras for navigation, and the video feeds captured by these cameras may record footage of ground-based citizens and their private spaces. 
A recent survey in the US found that 88\% of the participants were concerned about delivery drones recording their footage and using it for marketing and advertising~\cite{covington:zebra:2020}.
Admittedly, the delivery drone sector is regulated by government oversight (\eg~the identities of drones are generally known), and the drones belong to large e-commerce companies with reputations to protect. Nevertheless, privacy remains a problem because citizens have no way to reason about how the footage is stored or used. 
Recent drone-based food and coffee delivery trials by Alphabet Wing in various Australian cities led to citizen concerns and a parliamentary report calling for oversight on privacy~\cite{wsj:wing-coffee:2019,wing-trial:aus-parliament-report:2019}. Similar privacy violations in related domains, namely Google Street View, resulted in lawsuits~\cite{cnn:street-view} and strict regulations being imposed~\cite{wiki:street-view}. 
European drone vendors recently formed Drones4Sec~\cite{drones4sec}, a body to define protection of personal data in the era of drones (among other objectives). \newtext{Laws proposed by various countries have suggested that data gatherers (\eg~drone operators) must incorporate suitable accountability mechanisms to protect citizen privacy~\cite{puttaswamy:isc:2012}. Indeed, large e-commerce companies are often asked by governments to explain how they use the data they collect~\cite{npr:dec2020}.}


In this paper, we propose a framework called \textbf{\textsc{\pd}} to protect citizen privacy in the presence of delivery drones. \pd\ aims to implement the vision of a \textit{virtual privacy dome}, centred around the citizen, that protects their privacy from the prying cameras of delivery drones. \pd\ detects a delivery drone (or multiple drones) that may be in the vicinity of a citizen and determines whether that citizen is captured in the field of view of the drone's camera. \pd\ also provides mechanisms for the citizen to determine whether the pictures/video captured by the drone's camera are suitably sanitized to protect privacy, \eg~that all faces captured in the video feed are blurred (as is done in Google Street View). From the perspective of a ground-based citizen, \pd\ simply requires the citizen to install an application on their mobile phone. This mobile application helps determine the citizen's location, which is then used to identify drones in the citizen's vicinity and start the workflow in \pd. 

\pd\ is designed to be integrated with a city-scale regulatory authority that oversees delivery drone operations. These regulatory authorities are region-specific, \eg~the Federal Aviation Authority in the US~\cite{faa:remote-id:2021}, the Civil Aviation Authority in the UK~\cite{uk:caa}, or the Directorate General of Civil Aviation in France and in India~\cite{digsky,digskyR1E1}. The regulatory authority must be aware of the identity and current location of each delivery drone operating in the city. This assumption is realistic for the delivery drone sector and such requirements have been proposed in the drone laws of various countries (\eg~USA~\cite{faa:remote-id:2021}, France~\cite{ffna:remote-id:2018}, EU~\cite{eu:dronerules:2019}, Switzerland~\cite{swiss:dronerules:2021}, and India~\cite{digskyR1E1}), and security vendors are beginning to offer tracking solutions that can be incorporated into drones~\cite{thales:scaleflyt,911security}.


\pd\ has two components: \pdmpc, which we describe next, and \pdros, an auditing framework for the citizen to determine if privacy is maintained in the recorded footage.
\textbf{\pdmpc} allows a citizen to identify delivery drones whose cameras have the citizen in their field of view (\sectref{sec:pdmpc}). \pdmpc\ uses secure multiparty computation (MPC) between the regulatory authority's server and the citizen's mobile phone to accomplish this goal \textit{without revealing the citizen's location}. Two-party MPC, which we use, enables a pair of mutually-distrusting parties to collaboratively compute a function without revealing anything besides the function output. \pdmpc\ encodes a geometric computation that incorporates each drone's location, its direction of motion, the specifics of the drone's camera hardware and the citizen's location. This geometric computation determines if the citizen appears anywhere in the camera's field of view. \pdmpc's use of MPC ensures that it is able to accomplish this goal without requiring the regulatory authority to reveal the locations of all the drones in the city to the citizen.
 


However, \naive use of MPC in this setting presents scalability problems. Regulatory authorities operate at a city-scale, possibly tracking hundreds of delivery drones at any given time. Several thousand citizens may also simultaneously query the regulatory authority to identify drones in their vicinity. Off-the-shelf MPC protocols are computationally expensive and will have difficulty operating with the number of drones that one may expect at a city-scale. Moreover, each citizen communicates with the regulatory authority with their mobile phone, and traditional MPC protocols consume significant network bandwidth. \pdmpc\ is carefully engineered to scale to hundreds of drones, making it suitable for near-term city-scale deployments. Our evaluation of \pdmpc\ shows that each query from the citizen (to check their privacy at city-scale) only consumes up to \tabdataref{6.59MB} of mobile data for city-scale deployments of up to a thousand delivery drones. Note that the regulatory authority can execute queries from different citizens in parallel, and therefore automatically scales to an arbitrary number of citizen queries using cloud-based replication. 

Once a citizen identifies a drone(s) that has captured their footage, they may wish to ensure that their privacy is maintained in the recorded footage. This requires communicating with the drone, either directly or via the regulatory authority. A well-intentioned delivery drone must then convince the citizen that elements in the video that may violate the citizen's privacy have been sanitized appropriately. However, the citizen must have mechanisms to trust the drone's assertions that the data is sanitized. 

\textbf{\pdros} is an exemplar framework aimed to provide such assurances on ROS2-based drones (\sectref{sec:pdros}). ROS2---the Robot Operating System, version~2~\cite{ros2}--- is a popular middleware used in the software stacks of drones by numerous vendors. \pdros\ enhances ROS2 to audit data flows between applications within the drone. \pdros relies on trusted hardware on the delivery drone to provide the citizen with an audit trail of how their data is used within the drone. Many regulatory bodies do require delivery drones to be equipped with such trusted hardware (see \sectref{sec:overview} for references). 


Prior related methods in this area have been tailored toward ground-controlled drones~\cite{nassi:oakland:19,nassi:sp21,nassi:cscml21,birnbach:ndss:17}. They aim to  detect if a the drone has captured a citizen's footage in the first-person view exported to a ground-based human operator. \pd's methods are agnostic to whether the drone is ground-controlled or fully-autonomous. Companies such as Amazon are considering using fully-autonomous drones for their delivery fleets~\cite{amazon:prime:firstdelivery}, and prior methods will not work with such drones. \pd\ is the first system with mechanisms for citizens to obtain an audit trail from a drone that has captured their footage to maintain their privacy. 


To summarize, \pd's key contributions are:
\begin{mybullet}
\item \pdmpc, an MPC-based system for citizens to identify drones that have captured their footage while preserving location privacy of citizens. \pdmpc\ is carefully engineered to minimize network communication on citizens' mobile devices and work with city-scale drone deployments. Our evaluation simulating various city-scale deployments shows that for near-term deployments ($\sim$1000 drones), \pdmpc\ consumes lesser network bandwidth on the citizen's mobile device than streaming low-resolution YouTube videos.
\item \pdros, an auditing framework on ROS2-based drones, for citizens to determine if their captured footage is suitably sanitized. \pdros' workflow consumes a modest amount of additional CPU resources and power when implemented on a NVidia Jetson Xavier NX development board, with hardware similar to those on drones. 
\end{mybullet}

\mysection{Overview and Threat Model}
\label{sec:overview}


\begin{figure}[t!]
\centering
\includegraphics[width=0.95\linewidth]{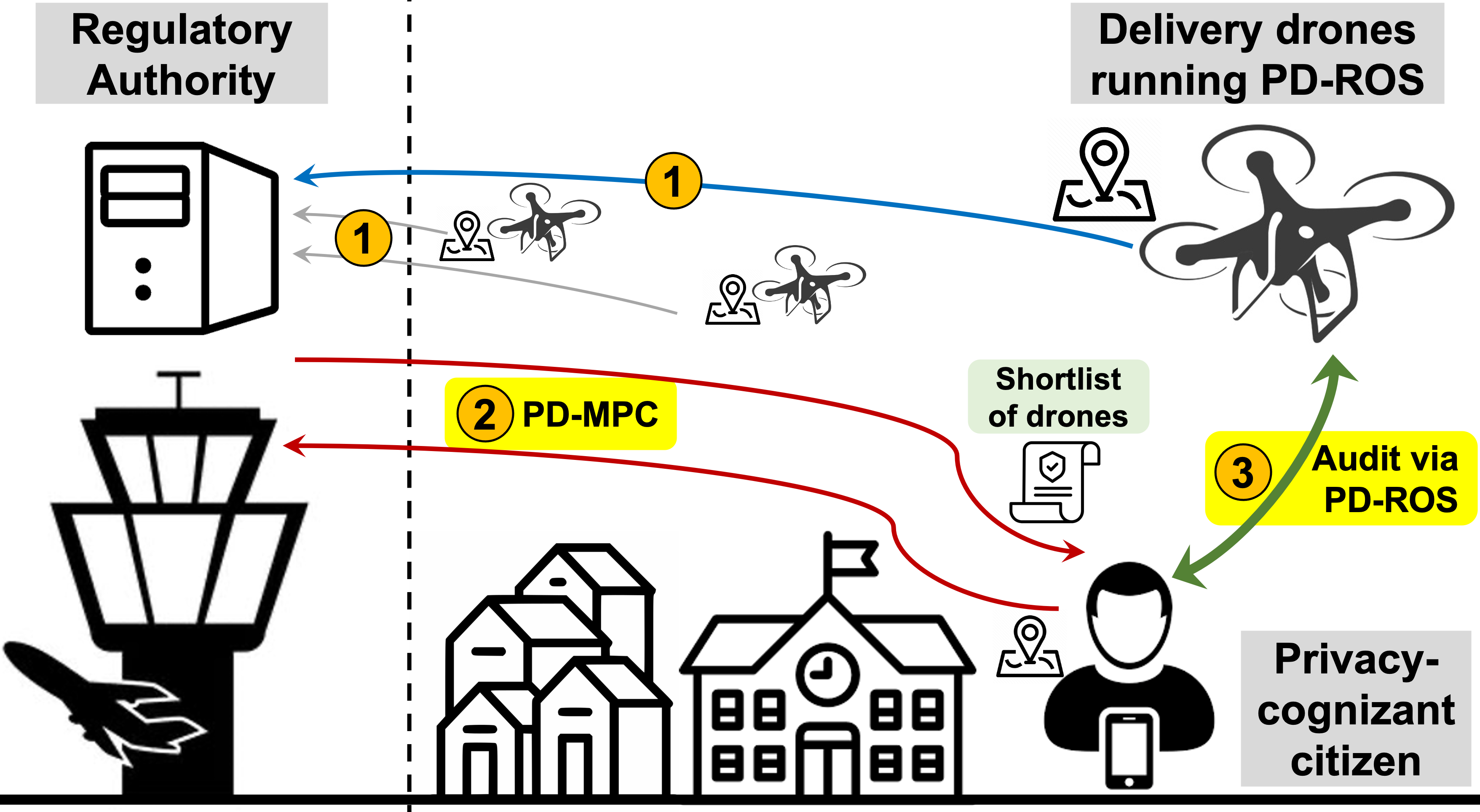}
{\small 
\begin{tabular}{p{0.98\linewidth}}
\rowcolor{LightGray}
\bfcircled{1}~Delivery drones register with a city-scale regulatory authority that oversees their operation. Each drone has a unique identifier and keeps the regulatory authority updated with its location.
\bfcircled{2}~Privacy-conscious citizen queries the regulatory authority with their current location (using \pdmpc), and obtains a shortlist of drones whose cameras may have the citizen in their field of view.
\bfcircled{3}~Citizen communicates with the drone, either directly or via the regulatory authority (without revealing their identity). Well-intentioned drones running \pdros\ can provide an audit trail to citizens showing that their privacy is maintained in the footage.
\end{tabular}}
\negspaceqtr
\negspace{-0.1cm}
\caption{Deployment vision for \pd.}
\label{figure:overallsetup}
\negspacetqtr
\negspace{-0.1cm}
\end{figure}

\figref{figure:overallsetup} presents the envisioned deployment scenario and overall workflow of \pd. \pd is designed to be integrated with a wide-area (\eg~city-scale) regulatory authority with oversight on delivery drone operations. We assume that the identity of each drone is known to the regulatory authority and that the drone provides real-time location updates to the regulatory authority (\textbf{Step \bfcircled{1}} in \figref{figure:overallsetup}). Indeed, such requirements have been stated in the drone laws of various countries. For instance, the US Federal Aviation Authority (FAA) has proposed that all delivery drone operations from 2023 must comply with remote identification rules, which require each drone to update the FAA with its identity, current location, altitude and velocity, in addition to other required data~\cite{faa:remote-id:2021}. France~\cite{ffna:remote-id:2018}, the European Union~\cite{eu:dronerules:2019}, Switzerland~\cite{swiss:dronerules:2021}, Australia~\cite{australia:oaic,wing-trial:aus-parliament-report:2019} and India~\cite{digskyR1E1} have also proposed similar regulations for drone operations. Some countries (\eg~India~\cite{digskyR1E1}) even require drones to declare their flight path and seek permission from the aviation authority prior to take-off. Security vendors have begun to offer commercial tracking solutions that can be incorporated into drones so that they comply with these regulations~\cite{thales:scaleflyt}. 

\pd\ additionally requires that the particulars of the drone's camera system (its focal length and sensor size) be known to the regulatory authority. This information is used in the geometric computation in \pdmpc. Although current regulations do not explicitly state that drones must make these details known to the regulatory authority, we feel that this is a reasonable assumption. These details are generally publicly available and associated with the drone's make and model.

When a citizen wishes to determine if a drone is recording their footage, they query the regulatory authority using an application on their mobile phone (\textbf{Step \bfcircled{2}} in \figref{figure:overallsetup}). The application can be configured to send such queries periodically. This application allows the user to specify a \textit{vicinity radius} centred around them. The regulatory authority will detect drones within the vicinity radius of the citizen that have captured the citizen's footage. Our focus in this paper is on the privacy of the citizen's footage itself. However, this application can easily be extended so that the citizen can also specify other locations (\eg~their home or yard) whose privacy they would like to protect in the footage recorded by delivery drones.

In \pd, citizens interact with the regulatory authority using \pdmpc, which uses a two-party MPC protocol. Its inputs are the details of all the drones in the city (the location, direction of motion, and camera particulars of each drone), as provided by the regulatory authority, and the citizen's location. \pdmpc\ uses this information to output to the citizen the identities of drones that may have captured the citizen's footage. \pdmpc\ goes beyond the traditional step of simply identifying drones in the citizen's vicinity (in fact, some drone-installed trackers already wirelessly broadcast their presence to nearby devices~\cite{911security}). \pdmpc\ uses the direction of motion of each drone and the details of the drone's camera in a geometric computation to determine whether the citizen's location is within the field of view of the drone's camera. Only such drones are shortlisted in the output of \pdmpc. As delivery drone operations begin to increase in popularity, we can expect several tens of drones to be present within typical vicinity radii that citizens may specify (\eg~500 metres). By shortlisting drones that have the citizen in their field of view, \pdmpc\ reduces the number of drones from which the citizen must request an audit trail in Step~\bfcircled{3}.

\pdmpc\ uses MPC to accomplish this goal without revealing the citizen's location to the regulatory authority. This is important because even though the regulatory authority is likely to be a trustworthy body, citizens will find it unpalatable to reveal their locations. Even if the citizen's queries anonymously reveal location, \ie~without disclosing the citizen's identity, prior work shows that citizens can be de-identified using historical query data (\eg~\cite{pedinti:ubicomp11}). 

\pdmpc\ encodes the entire aforementioned geometric computation using the machinery of MPC, thereby ensuring the citizen's location privacy. In that sense, \pd\ views the regulatory authority as an honest-but-curious participant. Delivery drone operators may also not wish to make the location of all their delivery drones publicly available to anybody that poses a query to the regulatory authority. \pdmpc\ has the pleasant side-effect of also ensuring this goal (to an extent), but this is a not our primary goal, whereas protecting citizen privacy is. The identities of drones that are in the citizen's vicinity are revealed to the citizen.


With the shortlist of drones in hand, the citizen can then request each delivery drone to show that it is taking measures to ensure privacy in the recorded footage (\textbf{Step~\bfcircled{3}}). This communication between the citizen and the drone can either happen directly, assuming supporting infrastructure for such direct communication exists, or mediated by the regulatory authority. This communication happens anonymously, without revealing the citizen's identity either to the drone or the regulatory authority. Each drone can then provide suitable proof to show that it is complying with region-specific privacy laws. For example, it could present an audit trail to the citizen showing that the footage is being sanitized on-board the drone or that the footage is being recorded at low resolution. 

Modern drones routinely capture video footage that is used to detect and avoid obstacles during navigation. Thus, it is impossible to offer a solution in which sensitive objects or people do not get captured in the video footage.  Prior solutions~\cite{nassi:oakland:19} that simply detect whether a citizen/object appears in the video footage will therefore flag (as suspicious) drones that may otherwise be well-intentioned, and happened to capture the footage of the citizen in their field of view during routine navigation. 
\pd's approach is to provide a framework (namely, \pdros) that allows well-intentioned drones to provide an audit trail to citizens that their privacy is being preserved in the video footage.  Drones that cannot provide a satisfactory audit trail or fail to communicate with the citizen can be reported (identities of shortlisted delivery drones are known to the citizen and the regulatory authority). \pd\ can thus abate citizen privacy concerns, such as those raised after a recent Alphabet Wing drone delivery trial~\cite{wsj:wing-coffee:2019}, which in turn incentivizes drone delivery companies to adopt a \pd-like solution.

\pdros is an auditing framework for ROS2-based drones. We chose to demonstrate our approach on ROS2 because of its popularity among drone vendors, \eg~various models sold by DJI, 3DR, Parrot, Gaitech, Erle, BitCraze, and Skybotix use ROS2. It should be possible to develop \pdros-like solutions for other drone software stacks as well. In our \pdros\ prototype, we ensure privacy by checking that applications on the drone only consume video footage from the camera after it has been sanitized, \eg~to blur faces that appear in each frame. ROS2 is a publish/subscribe system, in which applications publish and subscribe to \textit{topics}, \eg~the camera may publish to a topic called \rostopic{VideoFeed}, to which the navigation application may subscribe.  ROS2 sets up communication between applications by matching topics, and applications declare the topics to which they publish or subscribe in a manifest. Audit trails record the manifest of the application when it is launched. Manifests show how applications are permitted to communicate. Citizens can use them to verify that raw video footage is sanitized for privacy before being consumed by downstream applications. 


The key challenge, however, is to provide a basis for the citizen to establish trust in the integrity of the audit trail presented by a delivery drone. In \pd, we address this problem by requiring delivery drones to be equipped with trusted hardware. We use the ARM TrustZone~\cite{arm2009security} in our experimental prototype, but any similar attestation hardware should suffice. \pdros\ uses the ARM TrustZone to: \bfcircled{1}~run the trusted software that sanitizes image feeds;  \bfcircled{2}~procure and securely store the data producer/consumer information (the audit trail); and \bfcircled{3}~digitally signs and sends the audit trail to a citizen that requests a proof of footage sanitization.

It is natural to ask whether requiring delivery drones to be equipped with trusted hardware overly restricts the scope of \pd. In response, we note that drone laws are beginning to recognize and provide special certifications to drones that are equipped with trusted hardware. For example, India offers the higher-grade ``Level~1'' certification only to drones that have a hardware-backed trusted execution environment~\cite{digskyR1E1}. Drone vendors are also beginning to incorporate trusted hardware capable of securely storing information and performing cryptographic operations, \eg~the Wisekey Secure Element on the Parrot ANAFI Ai drone~\cite{parrot:anafi-ai}. 

That said, laws governing delivery drone operations are still evolving in various countries, and it is unclear if all countries will require delivery drones to be equipped with trusted hardware. For example, they may simply require that the drone be equipped with a certified software stack, but not require any hardware root of trust on the drone (\eg~the lower-grade ``Level~0'' certification given to drones in India~\cite{digskyR1E1}). Because \pdros is a set of tools atop the drone software stack, we expect that these tools can also be used on (ROS2-based) drones that lack trusted hardware. However, in the absence of trusted hardware the guarantees provided will be correspondingly weaker. For example, the audit trail cannot be digitally signed by the hardware root-of-trust, and the citizen will have to trust that the software stack on the drone is untampered to establish the integrity of the audit trail.

\anonsect{\underline{Threat Model.}}
Because \pd's main goal is to protect citizen privacy, its threat model is citizen-centric. The regulatory authority is assumed to be honest-but-curious. It truthfully engages with citizens to identify drones that have their footage, but citizens do not have to reveal their location to the regulatory authority. The regulatory authority must be able to track the locations of all delivery drones registered with it.

\newtext{We assume that delivery drones are owned by large e-commerce companies with reputations to protect. However, \pd\ only trusts these e-commerce companies to the extent that it expects them to operate drones equipped with a trusted execution environment, such as the ARM TrustZone. It expects the drones' identities and details of their camera hardware to be registered with the regulatory authority.} The public key associated with the trusted hardware can itself serve as the drone's identifier. We assume the existence of supporting public-key infrastructure, \ie~a certifying authority that issues digital certificates for the public keys of registered drones; the regulatory authority can serve this role. For \pd, the trusted-computing base on the drone is just the trusted hardware, capable of securely storing the audit trail and attesting the software stack on the drone. 
\newtext{Note that e-commerce companies have historically relied on  decentralized delivery fleet management. They often engage the services of \textit{third-party delivery service fleet operators (DSPs)} (\eg~see~\cite{dsp:amazon:cbnc,dsp:amazon,dsp:flipkart,swiggy:et:2022}), who procure and operate the delivery vehicles. While we can reasonably assume that e-commerce giants have no overt intention to break local privacy laws, the same cannot be assumed of DSPs. Indeed, \pd's threat model \textit{does not trust DSPs}, but does assume that the e-commerce company requires DSPs to operate drones equipped with trusted hardware, which in turn can attest that the DSP has not tampered with the drone's software stack.}

\newtext{One could ask why \pd-like mechanism is needed at all, if the e-commerce companies are assumed to be benign in intent. The answer is that it is still important to have accountability mechanisms in place to ensure that the e-commerce companies are abiding with local privacy laws~\cite{puttaswamy:isc:2012}. Having accountability mechanisms such as \pd\ both acts (1)~as a deterrent to the drone operators from violating local laws; and (2) as a confidence building means for citizens, who in turn will be more open to adopting delivery drone-based services. Indeed, history has shown that e-commerce giants are sometimes caught violating privacy of their clients~\cite{cnn:street-view,wiki:street-view,npr:dec2020,meta:reuters:2022}. This therefore underscores the need for a deterrent and an accountability mechanism available to citizens.
}

\pd's threat model excludes drones that are not registered with the regulatory authority and other rogue drones that deliberately try to hide their current location. We also exclude from our threat model drones that use attached cameras that are outside of the purview of the drone's attested software stack. For example, a rogue drone operator can attempt to bypass \pd\ by physically attaching a Go-Pro camera to the drone. Neither the trusted hardware on the drone nor the ROS2 software stack will have any control over such an attached camera. Other methods and regulations are required to deter such attacks. \newtext{We expect that e-commerce giants will not engage in deliberate attempts to violate local privacy laws because they have reputations to protect. However, individual DSPs may engage in such attacks. E-commerce giants can deter DSPs from engaging in such attacks by requiring the DSP to provide a live photograph of the drone prior to take-off from the warehouse, which can subsequently be examined for the presence of unauthorized attachments.}

\mysection{Shortlisting Drones with \pdmpc}
\label{sec:pdmpc}


\pdmpc\ is a system based on two-party MPC, involving a citizen and the regulatory authority. At the end of \pdmpc, the citizen has a short list of drones in whose field of view he/she appears. We first present the geometric computation used to detect if the citizen is in the drone camera's field of view (\sectref{sec:pdmpc:algo}), describe how to encode this computation in an MPC framework (\sectref{sec:pdmpc:mpc}), present implementation details (\sectref{sec:pdmpc:mpcimpl}) and an evaluation of our \pdmpc prototype (\sectref{sec:evaluation:pdmpc}).

\subsection{Drone Shortlisting Algorithm}
\label{sec:pdmpc:algo}

\begin{figure}[t!]
\centering
\includegraphics[width=0.85\linewidth]{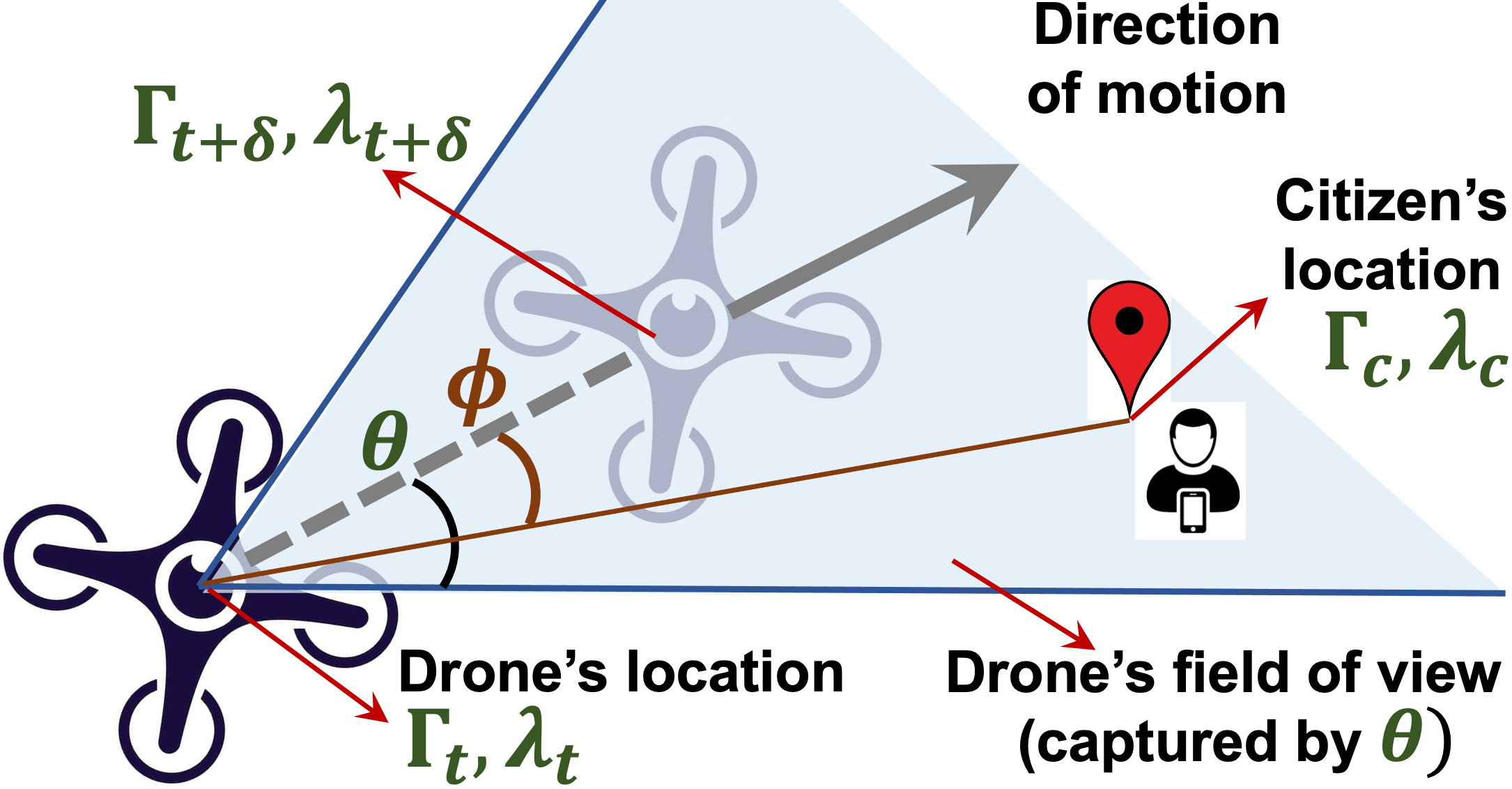}
{\small
\begin{tabular}{p{0.99\linewidth}}
\rowcolor{LightGray}
We use the drone's current GPS location ($\latitude_t$, $\longitude_t$) and location after a short time interval ($\latitude_\tplusd, \longitude_\tplusd$) to determine its direction of motion. The drone camera's field of view is abstracted by the parameter $\theta$. \pdmpc's geometric computation determines whether the citizen's location ($\latitude_c$, $\longitude_c$) is within the drone's field of view, \ie~whether $\phi<\theta$. For the field of view computation, \pdmpc\ converts location information captured as GPS coordinates into an equirectangular projection with 
($\latitude_t$, $\longitude_t$) as origin. \\
\end{tabular}}
\negspaceqtr
\caption{Setup for \pdmpc's geometric computation.}
\label{figure:pdmpc-geometric-computation}
\negspacehalf
\end{figure}

\figref{figure:pdmpc-geometric-computation} depicts the basic setup of the geometric computation. For simplicity, this figure considers a bird's eye view in two dimensions, with both the drone and the citizen on the same plane. Suppose we assume that a drone is within the citizen's vicinity radius, and that its location at a given instant in time is given by the pair of GPS coordinates ($\latitude_t$, $\longitude_t$). To determine whether the citizen at ($\latitude_c$, $\longitude_c$) is within its field of view, we need to know both the cone of vision of the drone's camera and the direction in which the drone is moving (assume for now a fixed, forward-facing camera, \newtext{which we subsequently relax}). The regulatory authority can compute the angle of the cone of vision ($\theta$ in \figref{figure:pdmpc-geometric-computation}) using the details of the drone's hardware with a well-known formula~\cite{mccollough:1893}: $\theta$ = $\arctan$($d$/$2f$), where $f$ is the focal length of the camera's lens and $d$ is the dimension of the sensor (\eg~the CCD sensor) used to digitally record the image. These parameters are typically associated with the make and model of the drone, which we assume is known to the regulatory authority. Algorithm~\ref{algorithm:fieldofviewdetection} therefore simply uses $\theta$ as one of the inputs.

The regulatory authority captures the drone's direction of motion using its GPS coordinates after a short interval of time, shown as ($\latitude_\tplusd$, $\longitude_\tplusd$) in \figref{figure:pdmpc-geometric-computation}. The citizen lies within the camera's field of view if the angle shown as $\phi$ in \figref{figure:pdmpc-geometric-computation} is less than $\theta$. On a two-dimensional plane with ($\latitude_t$, $\longitude_t$) as the origin, $\phi$ is the angle between the two vectors denoting the drone's direction of motion ($\vec{\mathbf{D}}$) and the citizen's location with respect to the origin ($\vec{\mathbf{C}}$). The value of $\phi$ can be determined using the dot product of $\vec{\mathbf{D}}$ and $\vec{\mathbf{C}}$, as shown in line~\mylineref{alg:fov:phi} of Algorithm~\ref{algorithm:fieldofviewdetection}. Recall that locations are reported as GPS coordinates, which are in latitudes and longitudes. To obtain the vectors $\vec{\mathbf{D}}$ and $\vec{\mathbf{C}}$, we need to obtain equirectangular projections of the GPS coordinates~\cite{snyder:1993} with ($\latitude_t$, $\longitude_t$) as the origin. \funcname{Vectorize}, called on lines~\mylineref{alg:fov:vec1} and~\mylineref{alg:fov:vec2} of Algorithm~\ref{algorithm:fieldofviewdetection} accomplishes this task.

\newcommand{\myskip}{}
\SetAlgoInsideSkip{myskip}
\begin{algorithm}[t!]
\SetAlgoLined
\LinesNumbered
\DontPrintSemicolon
\footnotesize
\KwIn{From citizen: $\latitude_c$, $\longitude_c$, \textsf{VicinityRadius}.}
\KwIn{From regulatory authority: 
        $\latitude_t$, $\longitude_t$, 
        $\latitude_{\tplusd}$, $\longitude_{\tplusd}$, $\theta$.}
\KwOut{\textsf{True} if citizen in field of view, else \textsf{False}.}
\tcp{\textrm{Each} ($\latitude$, $\longitude$) \textrm{is a GPS latitude/longitude pair.}}
\textsf{Dist} = \funcname{Distance}($\latitude_c$, $\longitude_c$, $\latitude_t$, $\longitude_t$)\; \label{alg:fov:distcalc}
\lIf{\upshape (\textsf{Dist} $>$ \textsf{VicinityRadius})}{\Return \textsf{False} \label{alg:fov:distthresh}} 
$\vec{\mathbf{D}}$ = \funcname{Vectorize}($\latitude_t$, $\longitude_t$, $\latitude_{\tplusd}$, $\longitude_{\tplusd}$) \; \label{alg:fov:vec1}
$\vec{\mathbf{C}}$ = \funcname{Vectorize}($\latitude_t$, $\longitude_t$, $\latitude_c$, $\longitude_c$) \; \label{alg:fov:vec2}
$\phi$ = $\arccos$(($\vec{\mathbf{D}} \cdot \vec{\mathbf{C}}$) / 
                   ($\lvert\vec{\mathbf{D}}\rvert$ $\times$ $\lvert\vec{\mathbf{C}}\rvert$)) \; \label{alg:fov:phi}
\lIf{\upshape ($\phi$ $\leq$ $\theta$)}{\Return \textsf{True} \textbf{else} \Return \textsf{False} \label{alg:fov:phicheck}} 
\SetKwProg{myproc}{Procedure}{}{}
\myproc{\upshape{\funcname{Vectorize} (Snyder~\cite{snyder:1993})}}
{
    \KwIn{GPS coords of \textsf{Origin} \& \textsf{Target}: ($\latitude_{o}$, $\longitude_{o}$, $\latitude_{t}$, $\longitude_{t}$).}
    \KwOut{Vector denoting \textsf{Target} in an equirectangular projection with \textsf{Origin} as origin.}
%
    X = $\mathbf{R}$ $\times$ ($\longitude_{t}$ - $\longitude_{o}$) $\times$ $\cos{(\latitude_{o}\times\frac{\pi}{180})}$ \;
    Y = $\mathbf{R}$ $\times$ ($\latitude_{t}$ - $\latitude_{o}$) 
    \tcp*{$\mathbf{R}$ = \textrm{Earth's radius}}
    \Return the vector [X, Y]
}
\caption{\funcname{DetectFieldOfView}.}
\label{algorithm:fieldofviewdetection}
\end{algorithm}


A few observations about Algorithm~\ref{algorithm:fieldofviewdetection} are in order:
\begin{mybullet}
\item 
\textit{Vertical Field of View.} The algorithm only considers the horizontal field of view, and ignores the vertical field of view. A citizen lies within the vertical field of view of the drone if $\phi_v$ $<$ $\theta$, where $\phi_v$ = $\arctan$($h$/\textsf{Dist}), where $h$ is the altitude of the drone, and $\textsf{Dist}$ is the distance between the drone and the citizen, as calculated on line~\mylineref{alg:fov:distcalc}. Although this calculation is simple enough to include in the algorithm, we chose not to do so because of two reasons. First, it is reasonable assume that a citizen appears within the vertical field of view of a drone within the vicinity radius. The citizen will likely be out of view only if the drone is too close to the citizen, \eg~hovering right overhead, in which case $\phi_v$$>$$\theta$). For drones that are in such proximity, we assume that the citizen's footage would have been captured on the drone's approach path, and that the citizen would be interested in obtaining an audit trail from that drone anyway. \newtext{If the drone is equipped with an optical flow camera (usually downward facing) it will capture the citizen's footage even if it is hovering right overhead.}
Second, the vertical field of view computation involves a division ($h$/$\textsf{Dist}$) and a trigonometric ($\arctan$) operation, both of which are expensive to execute when the algorithm is encoded in MPC. Moreover, as we discuss in \sectref{sec:pdmpc:mpc}, even precisely computing \textsf{Dist} is computationally expensive in MPC.

%
\item \textit{Gimbal-mounted Cameras/Multiple Cameras.}
The algorithm assumed a fixed camera, facing forward in the direction of motion of the drone. On drones with a gimbal-mounted camera, the camera may not necessarily point forwards in the direction of the drone's motion. Algorithm~\ref{algorithm:fieldofviewdetection} can easily incorporate drones with such cameras. The drone will have to additionally communicate the pitch and yaw of the gimbal to the regulatory authority. Using this information, the regulatory authority can calculate the angle ($\alpha$) between the drone's direction of motion and the camera's orientation. The angle $\alpha$ can be applied as a corrective factor to $\phi$ on line~\mylineref{alg:fov:phi}. The same idea can also be applied to account for the yaw and pitch of the drone itself, and suitable corrective factors can be applied to $\phi$. \newtext{The algorithm can also accommodate drones that are equipped with multiple cameras by considering a consolidated cone of vision that encompasses the cones of vision of each individual camera.}

\item \textit{Camera Resolution.}
The algorithm requires the citizen to provide a vicinity radius as input; only drones within this vicinity radius are considered. However, cameras differ in their resolution, with higher-resolution cameras capable of capturing sharp images from afar. Using a single vicinity radius ignores information about camera resolution. However, Algorithm~\ref{algorithm:fieldofviewdetection} can incorporate this information if the regulatory authority itself suggests suitable vicinity radii, based on the resolution of the drone's camera (which it can determine from the drone's make and model). In this case, citizens will specify the maximum acceptable resolution at which they are comfortable being captured in the footage, and the regulatory authority will suggest a suitable vicinity radius for each drone. 
\end{mybullet}



\subsection{Encoding as an MPC Computation}
\label{sec:pdmpc:mpc}

\pdmpc\ encodes \funcname{DetectFieldOfView}  as a secure two-party computation between the citizen and the regulatory authority. However, Algorithm~\ref{algorithm:fieldofviewdetection} requires a few modifications to make it secure and efficient for use in an MPC framework. We detail those modifications here, but first provide some background on MPC frameworks.

%
Broadly, there are two popular regimes of MPC protocols. One regime uses Yao-style garbled circuits (GC)~\cite{yao:focs86}. In GC, one party called the circuit garbler generates the garbled circuit and gives it to the other party who evaluates it. The circuit evaluator obtains the other party's inputs using oblivious transfer (OT) before evaluating the circuit, and the rest of the evaluation is non-interactive. The circuit evaluator then reveals the values at the output gate to the other party. In GC, fresh circuits must be used each time the two parties run the computation. 

A second regime uses secret-sharing, as exemplified by the GMW protocol~\cite{gmw:stoc87}. In this regime, both parties participate in the evaluation of the circuit. Each party holds its share of the value of a wire. During circuit evaluation both parties interact via a series of OT steps to exchange their wire shares to execute a gate in the circuit.  There is an extensive body of research in both regimes to optimize circuit evaluation, reduce the number of OT rounds, and to enable support for integer and floating point arithmetic and Boolean operations. Although our \pdmpc\ prototype is based on secret-sharing, this section discusses considerations for both regimes (\appref{sec:appendix:mpc-gc} discusses a GC-based implementation of \pdmpc). While garbling schemes aim to minimize interactive communication, the circuit produced by the circuit garbler is often quite large (we quantify this later in the section) and a fresh circuit must be provided to the circuit evaluator each time the circuit is evaluated. In contrast, secret-sharing based approaches minimize upfront network communication but require multiple interactions among the parties during the course of computation.

Algorithm~\ref{algorithm:fovdetection-mpc-oblivious} shows \funcname{DetectFieldOfView} adapted to MPC. We now describe the security and performance considerations that went into the design of Algorithm~\ref{algorithm:fovdetection-mpc-oblivious}.

\subsubsection{Security Considerations.} 
Note that Algorithm~\ref{algorithm:fieldofviewdetection} only considers inputs from \textit{one} drone. To protect citizen privacy, the algorithm needs to be executed with the coordinates of \textit{all} drones in the city. If Algorithm~\ref{algorithm:fieldofviewdetection} were na\"{i}vely iterated over all drones, it would be insecure, as described below.

First, note that the output must not be revealed to the regulatory authority because it could simply use the Boolean result of Algorithm~\ref{algorithm:fieldofviewdetection} to determine the set of drones in the vicinity of the citizen, and obtain an estimate of the citizen's location. We therefore modify the protocols to only reveal the output to the citizen. In secret-sharing MPC protocols, this goal is accomplished by ensuring that only the regulatory authority reveals its wire shares for the output gates to the citizen, but not vice-versa. In a GC-based setup, one could designate the citizen as the circuit garbler and the regulatory authority as the circuit evaluator. The citizen would modify the garbled tables for the output gates reveal only encoded values (seen by the regulatory authority), rather than the clear-text result of the computation. The encoded values can be decoded by the citizen, but not the regulatory authority. 

Second, even if the output were revealed only to the citizen, the nature of the computation in Algorithm~\ref{algorithm:fieldofviewdetection} has a subtle side-channel when executed on individual drones. While we can expect that there will be a few hundred drones at city-scale (in the near-term, scaling to a few thousand drones in 10-15 years), we only expect a small fraction of these to be within the vicinity radius of the citizen. For example, an Airbus study~\cite{airbus:blueprint:2018} estimated an average of 16,667 delivery drone flights per hour over Paris (roughly 100km$^2$) by 2035. This translates to under two drone flights per hour over a fixed 100m$^2$ area, assuming a uniform distribution.  As a result, we can expect that the more complex computation on lines~\mylineref{alg:fov:vec1} to~\mylineref{alg:fov:phicheck} will be triggered only for a few drones. In secret-sharing schemes, the side-channel manifests as network messages (\ie~the OT steps) exchanged during interactive execution. The regulatory authority can observe the number/size of messages exchanged during circuit evaluation and obtain an estimate of the number of drones in the vicinity of the querying citizen. In turn, this may be used to localize the citizen using the current density of drones in various locations of the city. In a GC-based setup, this side-channel would instead manifest as timing differences: the circuit would take longer to execute for drones in the vicinity of the citizen. If the regulatory authority is the circuit evaluator, it can similarly use this channel to identify drones near the citizen. 

To mitigate the side-channel, we run the computation in bulk for all drones. That is, rather than running the algorithm for each drone, we modify its inputs so that the regulatory authority provides the coordinates for all drones in one go. The algorithm iterates the distance computation over all the drones. Since all this happens \textit{within the MPC computation} the regulatory authority is oblivious to the identity of the drones that are in the citizen's vicinity.

\begin{algorithm}[t!]
\SetAlgoLined
\SetNlSty{}{\MPCALGOLINEPREFIX}{}
\LinesNumbered
\DontPrintSemicolon
\footnotesize
\KwIn{From citizen: 
    $\latitude_c$, 
    $\longitude_c$, 
    \textsf{LatVicinity}, 
    \textsf{LongVicinity},
    citizen's masking factors (drawn from $\mathbb{R^+}$): c$\latitude_1$, c$\longitude_1$, $\ldots$, c$\latitude_n$, c$\longitude_n$.}
\KwIn{From regulatory authority, identity, position and $\vec{D}$ vector for all drones: 
    (\textsf{Id}$^i$, 
     $\latitude_t^i$, 
     $\longitude_t^i$, 
     $\vec{\mathbf{D}}^i$) 
     for all drones in city ($i$=$1$$\ldots$$n$),
    regulatory authority's masking factors (from $\mathbb{R^+}$): r$\latitude_1$, r$\longitude_1$, $\ldots$, r$\latitude_n$, r$\longitude_n$.} 
\KwIn{Publicly-known: $\theta_1$, $\ldots$, $\theta_n$ for all drones.}
\KwOut{\textit{Revealed to citizen:} 
$\langle\textsf{DotP}^i, \textsf{NormSquare}^i, \textsf{NearbyLat}^i, \textsf{NearbyLong}^i\rangle$ for 
\underline{\textit{all}} drones.}
\textsf{[$\latitude$diff]} = \funcname{LatitudeDiff}($\latitude_t^1$, $\ldots$, $\latitude_t^n$,  $\latitude_c$)\; \label{alg:fovmpc:latdiff}
\textsf{[$\longitude$diff]} = \funcname{LongitudeDiff}($\longitude_t^1$, $\ldots$, $\longitude_t^n$, $\longitude_c$)\; \label{alg:fovmpc:longdiff}
\For{$i~\gets$~$1$~\KwTo~$n$}
{
    $\textsf{NearbyLat}^i$ = 
      (\textsf{$\latitude$diff[$i$]} - \textsf{LatVicinity}) 
        $\times$ c$\latitude_i$ 
        $\times$ r$\latitude_i$\; \label{alg:fovmpc:nearbylat}
    $\textsf{NearbyLong}^i$ = 
      (\textsf{$\longitude$diff[$i$]} - \textsf{LongVicinity}) 
      $\times$ c$\longitude_i$
      $\times$ r$\longitude_i$\; \label{alg:fovmpc:nearbylong}
    $\vec{\mathbf{C}}^i$ = \funcname{Vectorize}($\latitude_t^i$, $\longitude_t^i$, $\latitude_c$, $\longitude_c$)\; \label{alg:fovmpc:vecC}
      
    $\textsf{DotP}^i$ = $\vec{\mathbf{D}^i}\cdot\vec{\mathbf{C}^i}$\; 
    \label{alg:fovmpc:DotP}
    $\textsf{NormSquare}^i$ =  
          ($\lvert\mathbf{D}^i\rvert$)$^2$ $\times$ ($\lvert\mathbf{C}^i\rvert$)$^2$\; \label{alg:fovmpc:NormSquare}
    
    \textsf{Result}.add($\langle\textsf{Id}^i, \textsf{DotP}^i, \textsf{NormSquare}^i, \textsf{NearbyLat}^i,
    \textsf{NearbyLong}^i\rangle$)\; \label{alg:fovmpc:ResultAdd}
}
\Return \textsf{Result} (\textbf{\textit{revealed only to citizen}}) \label{alg:fovmpc:ReturnResulttoCitizen}

\par\vskip.5\baselineskip\hrule height 0.2pt\par\vskip.5\baselineskip

\tcp{\textrm{\textbf{The steps below happen on the citizen's phone (not as MPC)}}}

\For{\upshape $\langle\textsf{Id}^i, \textsf{DotP}^i, \textsf{NormSquare}^i, \textsf{NearbyLat}^i, \textsf{NearbyLong}^i\rangle$ $\in$ \textsf{Result} \label{alg:fovmpc:citizeniterate}}
{
    \If{\upshape ($\textsf{NearbyLat}^i$ $\leq$ 0) \textbf{and} 
                 ($\textsf{NearbyLong}^i$ $\leq$ 0) \label{alg:fovmpc:citizenIsnearby}}
    {
        $\phi_i$ = $\arccos$($\textsf{DotP}^i$/$\sqrt{\textsf{NormSquare}^i}$)\; \label{alg:fovmpc:citizenPhiI}
        \lIf{\upshape ($\phi_i$ $\leq$ $\theta_i$)}{shortlist the drone \textsf{Id}$^i$.} \label{alg:fovmpc:citizenShortListDrone}
    }
}
\caption{Adapting \funcname{DetectFieldOfView} to MPC.}
\label{algorithm:fovdetection-mpc-oblivious}
\end{algorithm}

However, running the computation in bulk for all drones alone is not sufficient for security. If lines~\mylineref{alg:fov:vec1}~to~\mylineref{alg:fov:phicheck} of Algorithm~\ref{algorithm:fieldofviewdetection} were executed conditionally only on drones in the citizen's vicinity, it exposes a subtle side-channel. In a secret-shared MPC setup, the choice of whether the computation within the conditional is executed (or not) would determine the number of OT messages exchanged between the regulatory authority and the citizen's mobile device. This, in turn, leaks information about the number of drones in the citizen's vicinity, which is undesirable. A similar side-channel would exist in a GC-based setup as well, in which the execution (or not) of the conditional would manifest as timing differences, observable by the regulatory authority, which evaluates the circuit. 

As a result, we designed Algorithm~\ref{algorithm:fovdetection-mpc-oblivious} to run the  steps to compute $\phi$
for \textit{all} drones, and not just those drones that are in the vicinity of the citizen. By executing the circuit on all drones in the city, Algorithm~\ref{algorithm:fovdetection-mpc-oblivious} makes the computation \textit{oblivious} to the number of drones in the citizen's vicinity. 

Our choice of using secret-shared MPC over garbled-circuit-based MPC was motivated by the size of the circuits needed to achieve the security properties discussed above. Although Algorithm~\ref{algorithm:fovdetection-mpc-oblivious} logically depicts the computation iterating over all the drones using a loop, the circuit represents the computation on lines~\mpclref{alg:fovmpc:nearbylat}~to~\mpclref{alg:fovmpc:ResultAdd} with the loop unrolled. Note that this is possible because the value of $n$ is predetermined, allowing the circuit to be generated for fixed values of $n$. In experiments with both secret-shared and GC-based MPC, we observed that the size of the circuits generated in a GC-based setup were 10-100$\times$ larger than the circuit sizes in a secret-shared setup. For example, for $n$=1000, the size of the circuit in our secret-shared MPC setup (using MOTION~\cite{motion:tops22}) is about \tabdataref{6.5MB}. In contrast, the size of the garbled circuit (generated using EMP~\cite{wang:emp:2016}, a popular GC framework) was \tabdataref{685MB}. Further, note that in a GC-based setup, a \textit{fresh} circuit must be used each time the two parties engage in computation, thus requiring the exchange of \tabdataref{685MB} \textit{each time} the citizen queries the regulatory authority, thus making the entire setup prohibitively expensive. 

\subsubsection{Performance Optimizations.} Our MPC encoding of Algorithm~\ref{algorithm:fieldofviewdetection} has four key performance optimizations:
\begin{mylist}
\item \textit{\funcname{Distance} computation.} Line~\mylineref{alg:fov:distcalc} of Algorithm~\ref{algorithm:fieldofviewdetection} computes the distance between the drone and the citizen. The Euclidean distance between the drone and the citizen, given their respective GPS coordinates, is computed using Haversine's formula~\cite{sinnott:1984}. Computing the Haversine formula involves trigonometric functions and square-root operations, which we wanted to avoid because they are well-known sources of inefficiencies in MPC algorithms~\cite{aly:eprint:2019}. For example, computing distance between a pair of points using Haversine's formula in 
MOTION~\cite{motion:tops22} requires a circuit with 284 gates and results in 179.9KB network traffic.
Vincenty's method~\cite{vincenty:1975} is more precise than the Haversine formula, but is even more expensive. 

We thus approximate the distance between the citizen and the drone using the difference between the GPS coordinates of the citizen and the drone. A difference of 0.001~degrees in the latitude values of two locations corresponds to a distance of 111~meters between them along the North-South axis~\cite{gpstodistance}. Likewise, the difference between their longitude values estimates their distance along the East-West axis (after suitably normalizing based on the latitude values, to account for an equirectangular projection). Subtraction operations can be implemented cheaply within MPC algorithms. Thus we modify our approach---instead of providing a Euclidean distance as a vicinity threshold, the citizen supplies a threshold of the difference between latitudes and longitudes. In our implementation, approximating the distance using this method requires a circuit with just 6 gates, and one distance computation costs just 0.239KB of network traffic. In Algorithm~\ref{algorithm:fovdetection-mpc-oblivious}, lines~\mpclref{alg:fovmpc:latdiff}-\mpclref{alg:fovmpc:longdiff} (\funcname{LatitudeDiff} and \funcname{LongitudeDiff}) compute the differences between the citizen's coordinates and the drone's coordinates in bulk, as an $n$$\times$1 matrix.

\item \textit{Lifting \funcname{Vectorize}.} Observe that the call to \funcname{Vectorize} on line~\mylineref{alg:fov:vec1} in Algorithm~\ref{algorithm:fieldofviewdetection} only uses inputs from the regulatory authority. This computation therefore does not have to execute as an MPC circuit. The regulatory authority can instead provide the $\vec{\mathbf{D}}$ vectors of drones as an input to the algorithm.

\item \textit{Early termination.} The computation on line~\mylineref{alg:fov:phi} of Algorithm~\ref{algorithm:fieldofviewdetection} requires inputs from both the citizen ($\vec{\mathbf{C}}$) and the regulatory authority ($\vec{\mathbf{D}}$) to compute the angle $\phi$. This calculation involves a division, a square root and an $\arccos$ operation. The square root appears as the final step in the calculation of the product of the L2-norms $\lvert\vec{\mathbf{D}}\rvert$$\times$$\lvert\vec{\mathbf{C}}\rvert$. Each of these operations is expensive in MPC.

We therefore modify the algorithm to omit the costly operations. The algorithm computes the value $\vec{\mathbf{D}}\cdot\vec{\mathbf{C}}$. It omits the square-root step in the computation of the norms within the MPC algorithm, leaving us with the values $\lvert\vec{\mathbf{D}}\rvert^2$ and $\lvert\vec{\mathbf{C}}\rvert^2$ (the square of the L2-norms), which we simply multiply. It reveals these values only to the citizen, who can then proceed with the square-root, division and $\arccos$ (in plaintext) on their mobile device to compute $\phi$ and shortlist drones accordingly. These steps appear as lines~\mpclref{alg:fovmpc:DotP}-\mpclref{alg:fovmpc:citizenShortListDrone} of Algorithm~\ref{algorithm:fovdetection-mpc-oblivious}. 
This approach results in considerable savings because the generated circuits are much smaller. In turn, this results in much reduced network communication overheads. 
For example, in our implementation 
of the computation in lines~\mylineref{alg:fov:phi} and~\mylineref{alg:fov:phicheck} (of Algorithm~\ref{algorithm:fieldofviewdetection}), 
early termination reduces the network traffic for \textit{one} iteration of the loop from \tabdataref{21.5MB} to \tabdataref{45.1KB}. This is because
in a secret-sharing MPC regime, early termination avoids additional OT steps that would otherwise be required if the computation happened within MPC. In a GC-based setup, early termination would reduce the communication from the garbler to the evaluator because of the reduction in the size of the circuit to be garbled.

There are two minor downsides to this approach. The first is that (unlike in Algorithm~\ref{algorithm:fieldofviewdetection}) the citizen learns the identities of all drones in their vicinity, and not just those of the drones that have captured their footage in their field of view. However, we feel that this is an acceptable tradeoff. Some drone-installed tracking devices already broadcast their their presence to nearby devices~\cite{911security} and the citizen is likely to be aware of the identities of drones close-by.

The second is that the value of $\theta$ associated with the drone must be available to the citizen to perform the comparison on line~\mylineref{alg:fov:phicheck} of Algorithm~\ref{algorithm:fieldofviewdetection}. We feel that the values of $\theta$ are not sensitive, and can be revealed publicly. In fact, the specification of most commercially-available drones is already available publicly, and the citizen would likely be able to compute the values of $\theta$ for most brands of drones. Revealing $\theta$ values only provides clues to the citizen about the brand of the delivery drone used, which we again feel is an acceptable tradeoff for performance. However, if delivery drone companies are hesitant to reveal $\theta$ values, we note that line~\mylineref{alg:fov:phicheck} alone can be encoded as a separate MPC computation, with the citizen supplying the $\phi$ values computed by the mobile application.

\item \textit{Avoiding Boolean comparisons in MPC.} Observe that the citizen performs the computations on lines~\mpclref{alg:fovmpc:citizenPhiI} and~\mpclref{alg:fovmpc:citizenShortListDrone} only on drones that are located within the vicinity threshold. The MPC computation can simply reveal a Boolean value for each drone, informing the citizen whether that drone is in the citizen's vicinity. For drone $i$ in Algorithm~\ref{algorithm:fovdetection-mpc-oblivious}, this Boolean is 
((\textsf{$\latitude$diff[$i$]} - \textsf{LatVicinity}) $\leq$ 0) $\wedge$
((\textsf{$\longitude$diff[$i$]} - \textsf{LongVicinity}) $\leq$ 0). Instead, Algorithm reveals the \textit{sign-preserving masked} values  $\textsf{NearbyLat}^i$ and $\textsf{NearbyLong}^i$ to the citizen, and the comparison is performed on the citizen's mobile device.\footnote{$\textsf{NearbyLat}^i$ and $\textsf{NearbyLong}^i$ are \textit{masked} values of (\textsf{$\latitude$diff[$i$]} - \textsf{LatVicinity}) and 
(\textsf{$\longitude$diff[$i$]} - \textsf{LongVicinity}), respectively, with the citizen and the regulatory each providing masking factors c$\latitude_i$, c$\longitude_i$, r$\latitude_i$, r$\longitude_i$ as additional inputs to the algorithm. The masking factors (drawn from $\mathbb{R^+}$) 
serve to protect the raw values of (\textsf{$\latitude$diff[$i$]} - \textsf{LatVicinity}) and (\textsf{$\longitude$diff[$i$]} - \textsf{LongVicinity}) from both the regulatory authority and the citizen, but \textit{preserve the sign of the result}. Masking is required because the raw values would reveal the locations of the drones to the citizen (or the location of the citizen to the regulatory authority). Note that \textsf{$\latitude$diff[$i$]} and \textsf{$\longitude$diff[$i$]} are bounded in range as they denote differences of latitudes and longitudes. Thus, the masking factors can be chosen from a large-enough range of enough range of the set of $\mathbb{R^+}$ numbers so that the multiplications on 
lines~\mpclref{alg:fovmpc:nearbylat} and~\mpclref{alg:fovmpc:nearbylong} are 
sign-preserving, \ie~no arithmetic overflows.}

We chose to reveal the values of $\textsf{NearbyLat}^i$ and $\textsf{NearbyLong}^i$ to the citizen to utilize optimizations implemented in the MOTION framework~\cite{motion:tops22}.
Observe that the MPC computation in Algorithm~\ref{algorithm:fovdetection-mpc-oblivious} is purely \textit{arithmetic} in nature (with only $+$, $-$ and $\times$ operators). MOTION uses algorithms tailored to make arithmetic computations fast, and introducing a comparison operator into the computation would require MOTION to switch to a Boolean representation of numbers to perform the computation. We observed that switching from the arithmetic to the Boolean world is an expensive operation in the MOTION framework. For example, with $n$=1000~drones, we observed that running Algorithm~\ref{algorithm:fovdetection-mpc-oblivious} consumes approximately \tabdataref{6.59MB}. In contrast, computing the Boolean value ((\textsf{$\latitude$diff[$i$]} - \textsf{LatVicinity}) $\leq$ 0) $\wedge$
((\textsf{$\longitude$diff[$i$]} - \textsf{LongVicinity}) $\leq$ 0) within MPC and revealing only the final result to the citizen consumes approximately \tabdataref{104.7MB}.
\end{mylist}


\subsection{Implementation}
\label{sec:pdmpc:mpcimpl}

As previously discussed, we used MOTION~\cite{motion:tops22} to implement 
Algorithm~\ref{algorithm:fovdetection-mpc-oblivious}. The regulatory authority reveals the value of $n$ in Algorithm~\ref{algorithm:fovdetection-mpc-oblivious} (the number of drones), and the circuit is generated for that value of $n$. Our experimental results indicate that our secret-sharing-based implementation scales to hundreds of drones at city-scale. With a city-scale deployment of a 1000 drones, each query consumes just \tabdataref{6.59MB} of mobile data. Assuming the citizen sends queries every 5 minutes, the bandwidth is \tabdataref{about half} of streaming a low-resolution video on YouTube (see \sectref{sec:evaluation:pdmpc}).

These results already are within the realm of practicality for near-term city-scale deployment of delivery drones. However, the bandwidth consumption and scalability of our approach can be improved further with a number of optimizations. Recent advances show that several steps of the OT can be completed as a pre-computation step, further reducing mobile data consumption~\cite{silver:crypto21}. Several MPC frameworks optimize mixed Boolean and arithmetic computations, and allow efficient switching between the Boolean and arithmetic worlds (\eg~\cite{aby:ndss15,aby2:usenix21}), which can also be explored. 


\subsection{Evaluation of \pdmpc}
\label{sec:evaluation:pdmpc}

We evaluated \pdmpc\ by studying how it performs under a simulated city-scale drone deployment. To model the regulatory authority server, we used a desktop with an Intel Core i7-7700 (3.60GHz) CPU, with 16GB RAM, running Linux 5.11.0-37 (Ubuntu 20.04). To model the citizen's mobile device, we used a NVidia Xavier NX board with (ARMv8.2 64-bit 6-core CPU, with 8GB RAM), running Linux for Tegra~\cite{l4t}. The mobile device garbles circuits while the server evaluates them.

\begin{figure}[t!]
\centering
\small
\begin{tabular}{c|c}
\thickhline
\rowcolor{LightGray}
\textbf{\scriptsize Number of drones in city}  & \textbf{\scriptsize Data consumed (MBs)}\\
\hline
\textbf{100}                             & 0.69\\
\rowcolor{LightGray}
\textbf{200}                             & 1.34\\       
\textbf{500}                             & 3.31\\
\rowcolor{LightGray}
\textbf{1000}                            & 6.59\\
\textbf{2000}                            & 13.14\\
\rowcolor{LightGray}
\textbf{10,000}                          & 65.60\\
\thickhline
\end{tabular}
\negspaceqtr
\caption{Per-query mobile data usage (MBs) on citizen's phone.}
\label{figure:mobile-data-usage}
\negspacetqtr
\end{figure}

\anonsect{Data Consumption on Citizen's Mobile Device.}
We measured the mobile data consumption on the citizen's device to pose queries to the regulatory authority. \figref{figure:mobile-data-usage} reports these measurements. The mobile data consumed depends the number of drones deployed in the city, which in turn determines the size of the circuit and therefore the OT messages that must be exchanged (\ie~the value of $n$ in Algorithm~\ref{algorithm:fovdetection-mpc-oblivious}). Thus, for our experiments we crafted inputs to our MPC implementation that vary the number of drones.

Observe from \figref{figure:mobile-data-usage} that mobile data usage increases as the number of drones in the city-wide deployment ($n$) increases. This is because the size of the circuit to increases proportionally with $n$, thereby resulting in a larger amount of data to be exchanged between the citizen's phone to the regulatory authority. The mobile data usage includes transmitting the circuit itself and OT during the computation. For a deployment with 1000 drones city-wide, each query from the citizen's phone consumes about \tabdataref{6.59MB} of mobile data. Assuming that the citizen sends queries every 5 minutes, this translates to approximately \tabdataref{79MB} of mobile data consumption an hour. This number is \tabdataref{about half} the network bandwidth required to stream a low-resolution video from YouTube (135MB/hour at 426x240p resolution, with a bitrate of 300Kbps). While there have been a number of trials of delivery drones, they are yet to be deployed at a large scale, and we expect the growth in this sector to be gradual. Considering, for example, the Airbus study cited earlier~\cite{airbus:blueprint:2018}, which projects 16,667 drones per hour over the city of Paris only by the year 2035, we feel that 1000 drones represents a conservative estimate of a near-term city-scale drone deployment. For a deployment with 10,000 drones city-wide, each query from the citizen's phone consumes about \tabdataref{65.6MB}. This shows that further improvements to MPC technology are needed to reduce this number and scale MPC city-wide as drone deployments increase.

\anonsect{Latency at Citizen's Mobile Device.}
We measured the end-to-end latency observed by a citizen from the time a query is issued to the time that the regulatory authority responds. We studied the latency both when the citizen and regulatory authority are on a 1Gbps LAN, and also on a slow WAN with highly variable upload speeds, in which the citizen's mobile device, connected on a 4G mobile data network (a mobile provider based in India), contacts the regulatory authority that runs on a virtual machine hosted on Azure (US East). \figref{figure:latency-citizen} reports the results of our experiments (average and standard deviation reported over 5 runs), which show that even in a slow WAN setting with $n$=1000, the citizen can obtain query results in approximately \tabdataref{16$\pm$4} seconds.

\begin{figure}[t!]
\centering
\small
\setlength{\tabcolsep}{4pt}
\begin{tabular}{c|r|r}
\thickhline
\rowcolor{LightGray}
\textbf{\scriptsize  $\downarrow$ Number of} & 
      \multicolumn{2}{c}{\scriptsize \textbf{\underline{Query latency at citizen's device}}}\\
\rowcolor{LightGray}
\textbf{\scriptsize drones in city} &    
      \lan\ (milliseconds)&
      \wan\ (seconds)\\
\hline
\textbf{100}       
          & 3.73$\pm$1.40
          & 4.52$\pm$1.29\\
\rowcolor{LightGray}
\textbf{200}       
          & 10.77$\pm$2.76
          &  8.84$\pm$1.59\\
\textbf{500}       
          & 28.36$\pm$7.63
          & 13.36$\pm$1.96\\
\rowcolor{LightGray}
\textbf{1000}      
          & 78.66$\pm$15.23
          & 16.33$\pm$4.09\\
\textbf{2000}       
          & 153.56$\pm$28.29
          &  23.82$\pm$4.06\\
\rowcolor{LightGray}
\textbf{10,000}      
          & 1281.32$\pm$126.20
          &   65.89$\pm$29.02\\
\thickhline
\end{tabular}
\begin{tabular}{p{0.98\linewidth}}
{\small 
Average RTT between the citizen and regulatory authority on our 1Gbps \lan\ is 0.220ms (measured with \texttt{ping}). In \wan, the citizen is connected via a 4G mobile data network, and the regulatory authority is an instance on Azure US East. The average RTT is 253.28ms, and the citizen's data upload speed during the experiment varied between 1.89Mbps to 12.96Mbps (averaging 6.78Mbps), measured with SpeedTest~\cite{speedtest}.}
\end{tabular}\\
\negspaceqtr
\negspace{-0.15cm}
\caption{Overall query latency at citizen's mobile device.}
\label{figure:latency-citizen}
\negspacetqtr
\end{figure}

Although our evaluation is citizen-centric, note that it also provides insight into the performance of the regulatory authority's server. Even in the most computationally- and communication-heavy case that we considered (10,000 drones city wide), the overall client latency was about \tabdataref{1.28s} in a LAN. The regulatory authority can parallelize MPC computation for different queries, therefore easily scaling up to an arbitrary number of querying citizens.

\anonsect{Accuracy of the \funcname{DetectFieldOfView} Algorithm.} Finally, we evaluated the precision of our approach at detecting whether an object is in a camera's field of view. For this evaluation, we conducted a field study in which we simulated the setup in \figref{figure:pdmpc-geometric-computation} using a fixed camera, \ie~we placed the the camera, denoting the drone camera, at specified location with a fixed orientation (\ie~the values of $\latitude_t$, $\longitude_t$ and the $\vec{\mathbf{D}}$ vector are known). We set up markers on the ground denoting the limits of the camera's cone of view, \ie~$\theta$ on either side of $\vec{\mathbf{D}}$.

We then repeated an experiment in which placed a person of interest, denoting the citizen, at a specified location (\ie~$\latitude_c$, $\longitude_c$ is known). Out of 20 trials of this experiment, we placed the person of interest within the camera's field of view in 10 trials, and outside the field of view in 10. \funcname{DetectFieldOfView} precisely determined that the person was within the field of view (or not) in 19 out of the 20 trials. The lone false positive was a case in which the person was close to the boundary of (and within) the field of vision, but was identified as being outside. We attribute this error to the quality of GPS measurements that we were able to obtain to determine location coordinates (we used the person's Samsung M31 device to determine their GPS coordinates). 
When we instead simulated the same experiment with Google Maps, using markers to identify the locations of the citizen, drone camera and the drone's direction of motion (using GPS coordinates from Google Maps), \funcname{DetectFieldOfView}'s accuracy was 100\%.

\mysection{Auditing Compliance with \pdros}
\label{sec:pdros}

With the shortlist of drones in hand, the citizen uses \pdros\ to obtain audit trails from drones and ensure that they are privacy-compliant. This section presents the goals and design space for \pdros (\sectref{sec:pdros:goals}), an overview of ROS2 (\sectref{sec:pdros:ros2}), a prototype implementation of \pdros\ atop ROS2 (\sectref{sec:pdros:main}), and its evaluation (\sectref{sec:evaluation:pdros}).

\subsection{Goals and Design Space Exploration}
\label{sec:pdros:goals}
\pdros\ is a set of tools that helps well-intentioned drones achieve the following goals:
\begin{mybullet}
\item (\goal{G1}) \textit{Use only sanitized data.} \pdros\ must ensure that camera data is never used unsanitized.
\item (\goal{G2}) \textit{Offer compliance proof.} \pdros\ must be able to convince a citizen that goal \goal{G1} is satisfied.
\end{mybullet}

To reliably offer proofs of compliance, \pdros\ assumes the presence of trusted hardware on the drone. This hardware must be capable of performing basic cryptographic operations, offer the ability to securely store an audit trail, and respond to citizen requests with a digitally-signed audit trail. Trusted hardware is typically endowed with a public/private key pair with the private key stored securely in the hardware, and the public key digitally certified by a certificate authority. This public key serves as the drone's identifier.

In our prototype, we use the ARM TrustZone~\cite{arm2009security}, which provides these features. An ARM TrustZone processor can be in one of two \textit{worlds} of execution at any given instant---a \textit{secure world} or a \textit{normal world}. The secure world, also called the trusted-execution environment (or the TEE), runs trusted software services. The normal world, or the rich-execution environment, runs untrusted applications and is normally the environment in which end-users of the device conduct the bulk of their activities. These features are implemented with traditional hardware-level protection, and a small, trusted, \textit{secure monitor} that executes at a higher processor privilege level than the OS in the secure and normal worlds. The two worlds interact via a secure monitor call (\texttt{smc} instruction) that allows world switches. The secure world implements features such as normal world attestation
(\eg~as in Samsung Knox~\cite{knox:ccs14}).

ARM TrustZone provides memory isolation for the secure world as a default feature, \ie~the normal world cannot map or access secure world memory. This feature provides us a path to achieve goal~\goal{G2} because audit trails and attestation reports can be safely stored untampered in the secure world. TrustZone also optionally offers the ability to securely split peripherals between the secure and the normal worlds, if the system-on-chip (SoC) includes the \textit{TrustZone Protection Controller} (TZPC)~\cite{arm:tzpc:nov04}. With TZPC, a peripheral can be assigned to the secure world for exclusive access, and therefore cannot be accessed by applications running within the normal world. Prior work~\cite{secloak:mobisys18} has used this to enable a trusted input path
via secure world control of certain peripherals (\eg~touchscreen).

On a TZPC-equipped SoC it is possible to accomplish goal~\goal{G1} as follows. The camera can be exclusively assigned to the secure world, and a trusted application executing in the secure world can sanitize the video feed before it is consumed by applications in the normal world. Any applications that require access to the raw video feed (\eg~a navigator that requires sharp video frames) would also execute within the secure world after \textit{a priori} vetting that they do not leak the raw footage. This would ensure that the camera data never leaves the secure world unsanitized, and the drone simply has to prove the existence of the above setup to a querying citizen, which can be accomplished by implementing secure boot and runtime attestation of the secure world. 

The precise notion of what it means to ``sanitize'' a video footage to preserve privacy is region-specific, and beyond the scope of \pd. In this paper, we intentionally do not commit to any particular method as an acceptable notion of video sanitization. For example, it could mean that the footage is obtained at low resolution. Or it could mean that sensitive objects, such as faces and car registration plates identified in the video feed, are identified and blurred (\ie~pixelated). This notion has been used in prior work~\cite{ipic:mobisys2016,privaros:ccs2020}, and is also the approach that is employed to preserve citizen privacy in Google Street View. In fact, prior work has even developed (MPC-based) methods to allow individual citizens to specify their own privacy policies, \eg~to have just their appearance blurred or edited out of the footage altogether~\cite{ipic:mobisys2016}. All of these are viable options within the broader \pd-framework, but for the sake of having a concrete policy for our discussion, we consider blurring frames (or \textit{all faces} within a frame) as our video sanitization policy. With TZPC, a trusted application that identifies and blurs faces in video feeds would achieve \pdros' goals.

We do not expect all drone platforms to have the TZPC on their SoC. For example, the experimental platform (a NVidia Xavier NX development board) that we use to build \pdros\ does not offer exclusive secure world access to peripherals. We therefore focus on how to achieve \pdros' goals even if the SoC lacks TZPC support. In the absence of TZPC, the video footage from the camera is accessible in the normal world. The mechanisms introduced by \pdros\ must therefore be tailored to the software environment executing in the normal world. We illustrate \pdros\ for drones running ROS2 in the normal world. We first provide with background on ROS2 and then describe the core components of \pdros.

\subsection{Background on ROS2}
\label{sec:pdros:ros2}

ROS2 is a popular middleware platform for robotics~\cite{ros2}. It is a set of libraries and user-space utilities that provide support for easy development of distributed robotics applications across a federation of robots. ROS2 offers a publish/subscribe model for robotics applications to communicate with each other. Applications publish messages labeled with specific topic names. Applications that subscribe to that topic can receive these messages from the publishing application. ROS2 uses the Data Distribution Service~\cite{dds,DDS_doc} to match pairs of applications that have such a publish/subscribe relationship. Each ROS2 application executes as a process atop the underlying OS, and ROS2 sets up either socket-based or shared-memory communication between a pair of applications.

ROS2 itself does not authenticate message senders, and all messages between applications are transmitted in the clear. This leads to a number of spoofing and eavesdropping attacks (\eg~\cite{dieber2016application,dieber2017security,mcclean2013preliminary,rodriguez2018message}). Moreover, any application can publish or subscribe to any topic. This leads to situations where a malicious application can publish a video feed under the same topic as a genuine application, and confuse downstream applications that consume the video feed. 
The community has therefore developed Secure ROS (SROS) to overcome these shortcomings~\cite{white2016sros,white2018procedurally,white2019sros1}. In SROS, TLS is used to secure the communication between applications. Further, each application must provide a manifest that declares the list of topics to which that application publishes or subscribes. The application's code and this manifest are then cryptographically bound via an X.509 certificate, signed by a trusted third party. SROS uses the X.509 certificate to detect and prevent the launch of any applications whose code or manifest have been modified. SROS also ensures that an application can only publish or subscribe to the topics explicitly identified in its manifest. \pdros\ builds upon ROS2 extended with SROS (we will use ``ROS2'' to refer to ROS2 enhanced with SROS). 

\subsection{Core Components of \pdros}
\label{sec:pdros:main}

On ARM TrustZone-enabled delivery drones, ROS2 and the corresponding application ecosystem execute in the normal world. The secure world runs a minimal trusted OS and trusted applications (the Trusty TEE~\cite{trusty} in our \pdros\ prototype). \pdros offers tools implemented in the normal world and in the secure world to restrict access to the raw video feed on drones in which the camera cannot be exclusively assigned to the secure world. It requires that raw access to camera hardware be restricted to a single ROS2 application. This requirement can easily be enforced by the normal world OS. 

\begin{figure}[t!]
\centering
\includegraphics[keepaspectratio=true,width=0.85\linewidth]{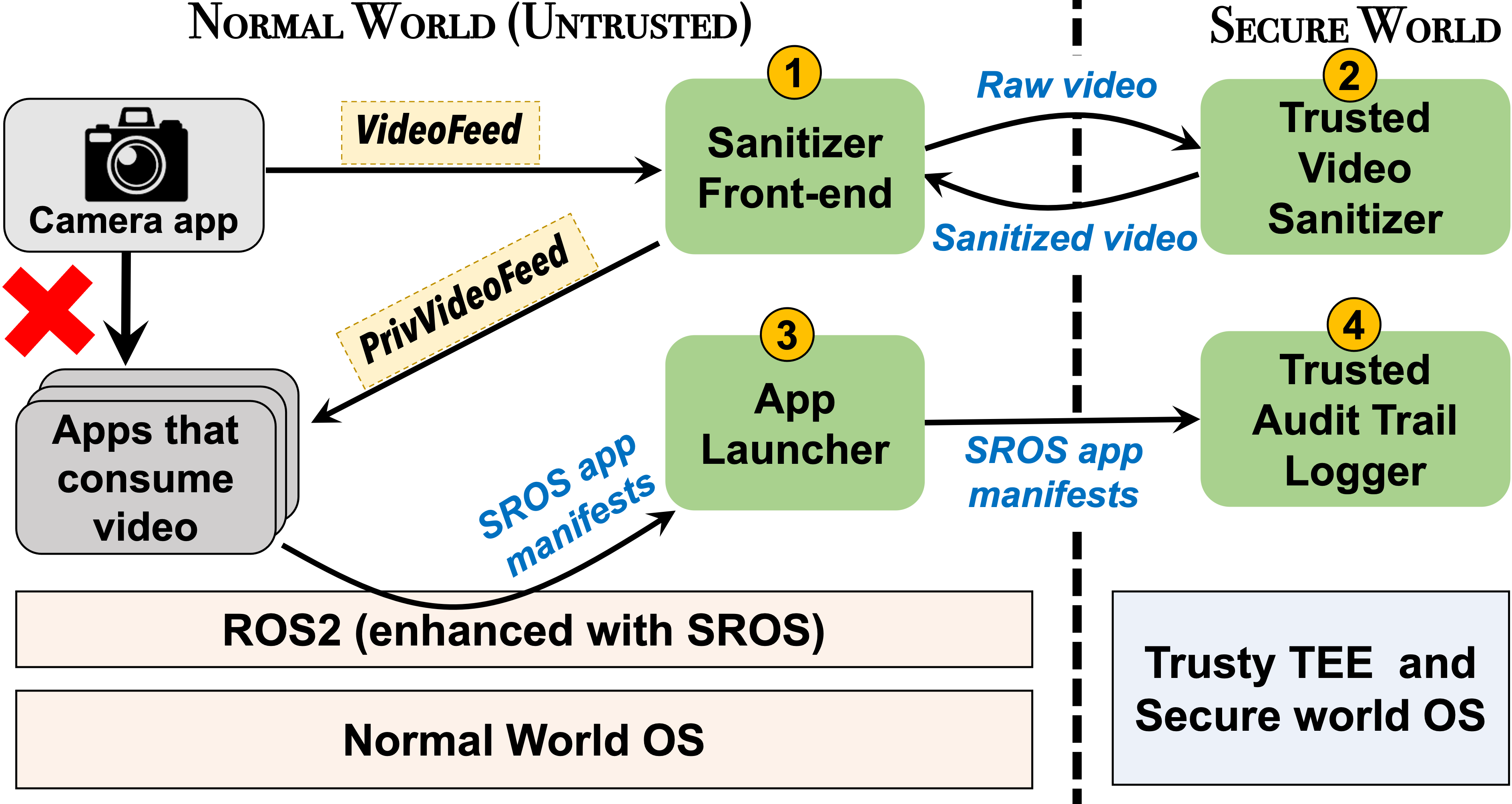}\\
{\small 
\begin{tabular}{p{0.98\linewidth}}
\rowcolor{LightGray}
\pdros\ provides well-intentioned drones tools to ensure that video footage is suitably sanitized, and to provide an audit trail to citizens to convince them so. \pdros\ introduces:
\bfcircled{1}~A ROS2 application that subscribes to the camera's feed (topic \rostopic{VideoFeed}). On a well-intentioned drone, no other applications will subscribe to this topic (they subscribe to \rostopic{PrivVideoFeed});
\bfcircled{2}~A trusted video sanitizer that runs in the secure world and applies region-specific privacy policies to the video feeds;
\bfcircled{3}~An agent in the normal world that collects the SROS manifests of applications that are launched for execution;
\bfcircled{4}~An audit logger that stores SROS manifests, and digitally signs and sends them in response to citizen queries.
\end{tabular}}
\negspaceqtr
\negspace{-0.1cm}
\caption{Setup of ARM TrustZone-based drone with \pdros.}
\label{figure:pdros-components}
\negspacetqtr
\negspace{-0.1cm}
\end{figure}

Suppose that this camera application publishes its feed to a topic called \rostopic{VideoFeed}. Downstream applications can consume the video feed by subscribing to this topic. \pdros\ requires that applications on well-intentioned drones subscribe instead to a topic called \rostopic{PrivVideoFeed}. \pdros\ provides a dedicated ROS2 application---the Sanitizer Front-End (Sanitizer-FE)---that exclusively subscribes to \rostopic{VideoFeed} and publishes to \rostopic{PrivVideoFeed} (see \figref{figure:pdros-components}). Sanitizer-FE uses the traditional publish/subscribe abstraction, thereby allowing other downstream ROS2 applications to interact with it without requiring any invasive changes to their code. At the back-end, Sanitizer-FE uses a secure monitor call (\texttt{smc}) to perform a world switch, and interact with a trusted video sanitizer in the secure world. This video sanitizer is entrusted with applying the region-specific privacy policy, \eg~blurring faces in the video feed. As discussed earlier, ROS2 applications that require raw access to \rostopic{VideoFeed} can be accommodated as exceptional cases after vetting that they do not intentionally leak the feed. 

This setup suffices to ensure goal \goal{G1} on a well-intentioned drone. However, the drone must also convince a citizen that this setup exists on the drone (goal \goal{G2}). To accomplish goal \goal{G2}, \pdros relies on the way application manifests work in SROS. As previously discussed, the manifest is cryptographically-bound to the application's identity, and SROS uses the X.509 certificate of the application to check that the manifest and the application's code have not been tampered. The manifest specifies the topics that the application publishes/subscribes to, and SROS ensures that the application does not deviate from this specification at runtime. 

\pdros\ extends ROS2's application launcher to store an application's manifest in the secure world when it is launched in the normal world. The trusted audit logger in the secure world attests the normal world kernel, ROS2 and SROS and stores the attestation report in the audit trail whenever an application is launched. Entries in the audit trail are time-stamped, and the citizen either requests the entire audit trail, or a snippet for a particular time interval. Upon a citizen query, the secure world digitally signs and sends the audit trail. The citizen then uses the audit trail as follows:
\begin{mylist}
\item \textit{Check normal world OS/ROS2/SROS integrity.} The citizen first uses the attestation report to verify the integrity of the normal world OS, ROS2, and SROS. This step is critical because \pdros\ relies on the normal world OS to ensure that access to the camera hardware is restricted to a single application. Because \pdros\ relies on SROS to enforce application manifests, the citizen must ensure the integrity of ROS2/SROS in the normal world.
\item \textit{Check integrity of Sanitizer-FE and ROS2 application launcher.} As is standard~\cite{tpm:sec2004}, we assume that the normal world OS also includes in the attestation report the integrity measurements of applications that it launches. The citizen uses these measurements to verify the integrity of Sanitizer-FE and the application launcher.
\item \textit{Check publish/subscribe patterns.} SROS application manifests limit how applications communicate. Thus, the citizen simply needs to verify that applications that execute in the normal world do not subscribe to any topics published by the camera application (\ie~\rostopic{VideoFeed}), and that Sanitizer-FE subscribes only to \rostopic{VideoFeed} and publishes to a single topic, \rostopic{PrivVideoFeed}. Together with the runtime assurance provided by SROS and the trusted video sanitizer in the secure world, the citizen can be assured that the video footage is being sanitized before use. While the topics specified in the manifest of an application represent the sandbox within which the application can operate, they may not always publish or subscribe to all the topics in the manifest. Although not in our prototype, \pdros\ could use an agent to dynamically query ROS2 to determine the publish/subscribe graph, and store that instead in the audit trail.
\end{mylist}
Citizens can either query the shortlisted drones directly (if such communication support exists), or via the regulatory authority. Such queries must obviously be anonymous to protect the citizen's identity, failing which the regulatory authority can simply use the set of queries from a citizen to compromise their location privacy. Fortunately, such a query interface can easily implemented using well-known methods (\eg~Tor).

The assurances provided by the audit trail rely on certain assumptions that are standard in hardware-based attestation. For example, we cannot defend against zero-day exploits on SROS, ROS2, the normal world kernel, or any of the components of \pdros.  We also assume that SROS completely mediates application communication within the normal world. A malicious ROS2 application can attempt to bypass SROS by directly invoking OS abstractions, \eg~via raw sockets, to communicate with a colluding application~\cite{privaros:ccs2020}. Prior work has attempted to harden the normal world kernel against zero-day exploits~\cite{knox:ccs14,sprobes:2014}, and to ensure that applications only communicate under the purview of SROS~\cite{privaros:ccs2020,ros+apparmor}. Those methods also apply to our setting. \appref{sec:appendix:pdros-analysis} provides a detailed analysis.

Given our goals, it is natural to ask whether audit trails produced by \pdros\ alone do not suffice to protect citizen privacy. The regulatory authority could itself periodically query all delivery drones in operation, check the audit trails to determine whether the drones are in compliance, and publish a compliance report for all drones on a public forum for all citizens to see. If the compliance report also includes the locations of drones (provided a drone operator permits the release of the location information for all its drones, city-wide), citizens can determine the compliance status of drones in their vicinity using this public forum. 

The key shortcoming with this deployment model is that does not provide any mechanism for the citizens to seek further accountability that is otherwise possible with \pdmpc. For example, with \pdmpc, the citizens know precisely which drones have captured their camera footage. This allows the possibility of an accountability mechanism via which citizens can request their footage from those drones and verify that it has been sanitized. This is akin to citizens being able to request their activity reports on platforms such as Google or Facebook, or citizens being able to view their footage on Google Street View and confirm that it has been sanitized.



\subsection{Evaluation of \pdros}
\label{sec:evaluation:pdros}

We implemented and evaluated \pdros\ on a Nvidia Xavier NX development kit, equipped with a 6-core Nvidia Carmel ARMv8.2 64-bit CPU and 8GB RAM. It also has a 384 core GPU with 48 tensor cores that makes it ideal for heavy video processing. This hardware is similar to that found as the companion boards on many commercial drones. We chose this development kit because it is equipped with the TrustZone and has an open and programmable secure world. We configured the secure world to run the Trusty TEE~\cite{trusty}, which allows for easy development of trusted applications. The normal world runs Linux for Tegra (L4T)~\cite{l4t}, which consists of the Linux kernel 4.9, a bootloader, and supporting drivers. We installed a ROS2-based environment in the normal world


\begin{figure}[t!]
\centering
\footnotesize
\includegraphics[width=0.6\linewidth]{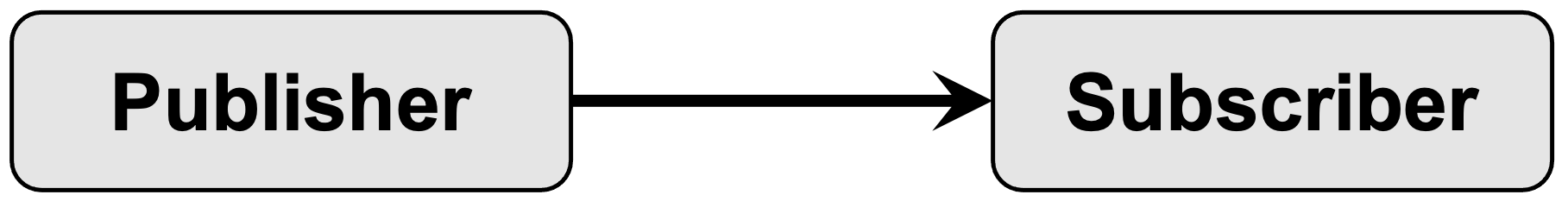}\\
\bfcircled{a}~\textbf{Baseline setup.}\\
{~}\\
\includegraphics[width=0.7\linewidth]{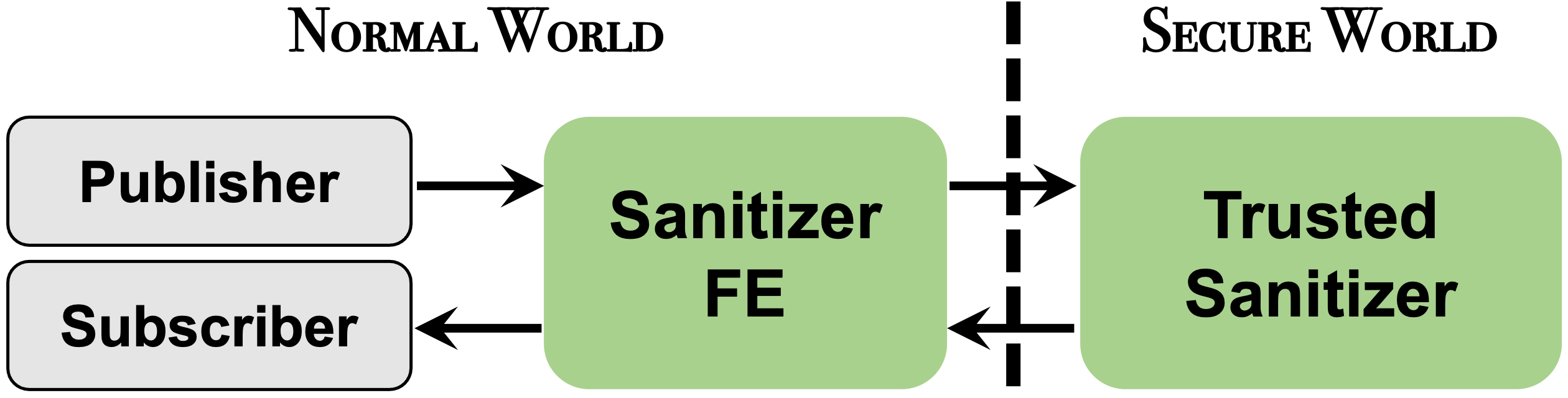}\\      
\bfcircled{b}~\textbf{Setup with redirection and blurring.}\\
%
\renewcommand{\arraystretch}{0.9}
\small
\begin{tabular}{lrr@{~~~~}r}
\thickhline
\rowcolor{LightGray}
\textbf{Metric}                
                & \textbf{Baseline} 
                & \multicolumn{2}{c}{\textbf{With redirection}}\\
\hline
\textbf{Time}            & 16.643s           & 16.687s & (+0.26\%)\\
\rowcolor{LightGray}
\textbf{CPU Utilization} & 18.68\%           & 31.82\% & (+70.34\%)\\
\textbf{Power Use:}         &                   && \\
\textit{$\bullet$ System 5V rail} 
                & 4138.58mW         & 4980.73mW & (+20.34\%)\\
\textit{$\bullet$ SoC rail}         
                & 1217.15mW         & 1250.11mW & (+2.71\%)\\
\textit{$\bullet$ CPU+GPU rail}     
                & 744.22mW          & 1502.87mW & (+101.8\%)\\
\thickhline
\multicolumn{4}{c}{\bfcircled{c}~\textbf{Performance impact of flow redirection on NVidia Xavier.}}
\end{tabular}
\negspaceqtr
\caption{\pdros\ experimental setup and results.}
\label{figure:pdros-experiments-results}
\label{figure:pdros-experiments-setup}
\negspacetqtr
\end{figure}

To evaluate the cost of sanitizing video feeds, we consider the setup shown in \figref{figure:pdros-experiments-setup}. We built a pair of ROS2 applications, the first of which publishes a video feed that the second consumes directly, as shown in \figref{figure:pdros-experiments-setup}\bfcircled{a}. This is our baseline, which we run with a workload in which the publisher sends a video feed consisting of 250 frames (resolution 320$\times$240 pixels, at 30 frames/second). The publisher converts each frame into ROS2's message format, which the subscriber receives and converts back to a video frame. We then implemented redirection and sanitization of the video feed as shown in \figref{figure:pdros-experiments-setup}\bfcircled{b}, in which Sanitizer-FE exclusively subscribes to the feed of the publisher, and redirects the flow to a trusted sanitizer in the secure world. The trusted sanitizer applies a simple box blur filter to the image. We intentionally kept the functionality of the trusted sanitizer simple to show the basic cost of redirection. More complex image processing logic can obviously be built within the trusted sanitizer, with a corresponding increase in resource consumption. The trusted sanitizer returns the modified feed to Sanitizer-FE, from which the subscriber application consumes it.

%
%
%
%

\figref{figure:pdros-experiments-results}\bfcircled{c} presents the results of our experiments. We measured the end-to-end latency of the application workflow (from publishing to receiving the video feed), the CPU utilization. We also used the in-built 3-channel INA3221 power monitor on the Xavier board to measure power consumption at the system, SoC and CPU+GPU rails, respectively. As our results show, the overall application latency on this lightly-loaded drone is minimally impacted, with the video being blurred in near real time. This operation only led to a modest increase in CPU utilization and power draw.

We also evaluated the overall cost of attesting the normal world (kernel, ROS2, and SROS), and storing application manifests in the secure world. Recall this step is performed when the application is launched, and thus increases launch time. We measured the time to launch the \texttt{cam2image} application, which is part of the standard ROS2 distribution. Without \pdros\ this application takes 4.78s to launch on our hardware platform, while it takes 10.66s with \pdros.


\mysection{Related Work}
\label{sec:relwork}

Although the focus of this paper is on citizen privacy in the presence of delivery drones, 
a number of prior works have considered security and privacy of drones and other 
aerial vehicles, broadly considered. Nassi~\etal~\cite{nassi:oakland21} provide a comprehensive overview of the security and privacy issues in the era of drones. 

Recent work has attempted to detect privacy violations committed by drones that are controlled by a ground-based operator equipped with a first-person view (FPV). These works leverage the observation that such a drone must wirelessly communicate with the operator to export the camera's view to the FPV. Wi-Fly~\cite{birnbach:ndss:17} aims to detect drones that hover outside the windows of homes. It uses a window-mounted sensor that detects a drone approaching the window by studying variations in the received signal strength (RSS) at the sensor. Nassi \etal's work~\cite{nassi:oakland:19,nassi:cscml21,nassi:sp21} detects whether an object (or person) is captured in the FPV. Their work is based on the observation that if an object is being recorded in the FPV, then physical perturbations of the object (\eg~shining light on it) manifest as observable changes in the encrypted wireless channel between the drone and the remote control. They use a ground-based detector to intercept this encrypted wireless channel, and then employ cryptanalysis techniques that leverage this observation to detect privacy-violating drones. 

However, this prior work suffers from three important shortcomings. First, the detection methods are \textit{tailored to drones that export an FPV} to a ground-based operator. They rely in a key way on the detector having access to the wireless channel that the drone uses to export the FPV. They are thus not applicable to autonomous drones, which may not export such an FPV or have a ground-based operator. 
Second, they \textit{do not offer an end-to-end solution} to a citizen who may wish to determine whether their privacy is violated. That is, while they may help detect that a citizen is captured in the FPV, they offer the citizen no way to deal with the violation or communicate with the drone to query how the recorded video feed is used or stored. Indeed, these methods are not in any way tied to any regulatory framework that offers the citizen to reason about how the captured data is used. Finally, these methods require \textit{active citizen participation}, either by installing detectors on their home windows (in the case of Wi-Fly~\cite{birnbach:ndss:17}) or physically perturbing the object/citizen suspected of being observed (in the case of Nassi \etal's work~\cite{nassi:oakland:19,nassi:sp21,nassi:cscml21}). However, some synergies with \pd\ exist. For example, a citizen can use Nassi \etal's methods to detect FPV-based drones that have him/her in their field of view (instead of \pdmpc), and then use \pdros to engage with the drone to obtain an audit trail. On the flipside, while \pd's approach is restricted to delivery drone operations that are overseen by a regulatory authority,  Wi-Fly~\cite{birnbach:ndss:17} and Nassi \etal's methods~\cite{nassi:cscml21,nassi:oakland:19,nassi:sp21} apply to any FPV-based drones.

While Wi-Fly and Nassi \etal's work focuses on how individual citizens are impacted, Privaros~\cite{privaros:ccs2020} and AliDrone~\cite{alidrone2018icdcs} develop methods to regulate delivery drones over well-demarcated host airspaces that may dictate that specific policies are to be obeyed within the airspace. For example, a college campus or an apartment complex may require the delivery drone to ensure that images and video recorded in the their airspace be blurred suitably, or that drones follow certain pre-identified drone lanes during their visits. Privaros develops mandatory access control extensions for ROS-based drones that can accept and enforce policies specified by ground-based hosts. Like \pdros, Privaros also relies on trusted hardware to prove to hosts that the delivery drones are in compliance. However, unlike \pd, Privaros does not focus on privacy of individual citizens, nor does it offer a method to determine whether a citizen's footage is captured by the drone. AliDrone~\cite{alidrone2018icdcs} is tailored to ensure that drones remain on drone lanes during their delivery runes. It uses trusted hardware to securely store proofs-of-alibi (GPS coordinates of the flight path) that can be used to prove to host airspaces that the drone was in compliance. PROTC~\cite{liu2017protc} also uses trusted hardware on drones with a focus on protecting the drone software stack and peripherals from malicious attacks and rootkits~\cite{liu2017protc}.

In contrast to the above works, which mainly focus on privacy, there is a significant body of work on drone security. Given the near-daily news stores about drones being used to conduct various terrorist attacks (\eg~Venezuela~\cite{koettl:nytimes:2018}, Iraq~\cite{iraq:nov21} and Japan~\cite{japan:apr15} being prominent examples), it is not surprising that much of the focus is on detecting unauthorized drones. Methods to detect drones range from the use of radar~\cite{eshel13}, Lidar~\cite{church:optics18,qsystems,3deo} radio-frequency~\cite{matthan:mobisys:2017}, computer vision~\cite{rozantsev:cvpr:2015}, and acoustic signatures~\cite{busset15,case:naecon08}. These methods can complement \pd\ in city-scale deployments to detect or deter drone flights outside of the purview of the regulatory authority. Security research focused on delivery drones has mainly considered reliable package delivery. Here, prior work has focused on methods to mutually authenticate delivery drones and intended recipients, for example, using the sound signature of the drone~\cite{sounduav:dronet19}. Researchers have also developed methods to ensure that delivery drones are not sabotaged in-flight, by developing methods to detect and avoid projectiles thrown at them~\cite{garg:hotmobile2020}.

Abidi \etal~\cite{abidi:ndss22} consider a setting in which citizens query aggregate pollution statistics collected from sensors fitted on taxi fleets. As in \pd, they protect citizen privacy using MPC. However, they also consider the dual problem of protecting the privacy of the taxi fleet using differential privacy, \eg~to hide the distribution of taxis of one fleet operator from a competitor. While Abidi \etal's work is in a different domain, \pd\ can borrow similar ideas to also protect drone fleet privacy from competing fleet operators.

\mysection{Conclusion}
\label{sec:conclusion}

Even as e-commerce companies work on the technology and infrastructure to enable drone-based delivery, citizens are concerned about on how their privacy will be impacted once it sees wide deployment. Clearly, much needs to be done to abate these concerns, in the form of governmental regulation and enforcement methods. \pd\ is a step in that direction. It describes a framework that can be integrated with city-scale regulatory authorities that oversee drone operations. The framework allows citizens equipped with a mobile phone to determine if their footage is recorded by drones in their vicinity, and then request an audit trail from those drones to determine if they comply with region-specific privacy laws. \pd\ accomplishes all these goals without revealing the citizen's location, and while only consuming mobile data comparable to streaming low-resolution videos. Our experiments indicate that \pd\ can to scale to near-term city-scale drone deployments. \pd's auditing mechanisms impose only a modest runtime overhead on CPU utilization and power consumption on the drone.

\indent\par
\anonsect{Code availability.} 
Available at \href{http://dx.doi.org/10.5281/zenodo.6442206}{\texttt{DOI: 10.5281/zenodo.6442206}}.


\newpage
\appendix
\section*{APPENDIX}
\section{\pdmpc using Garbled Circuits}
\label{sec:appendix:mpc-gc}

It is natural to ask whether Algorithm~\ref{algorithm:fieldofviewdetection} can be encoded in MPC using garbled ciruits. As already mentioned in \sectref{sec:pdmpc:mpc}, we observed that the sizes of the circuit denoting the MPC encoding in  Algorithm~\ref{algorithm:fovdetection-mpc-oblivious} using a GC-based setup were 10-100$\times$ larger than the circuit sizes in a secret-shared setup. For $n$=1000, the size of the secret-shared setup using MOTION~\cite{motion:tops22}) is about \tabdataref{7MB}. In contrast, the size of the garbled circuit using EMP~\cite{wang:emp:2016} was \tabdataref{685MB}. Thus, the MPC encoding shown in Algorithm~\ref{algorithm:fovdetection-mpc-oblivious} would be prohibitively expensive if implemented using garbled circuits. 

The key problem that makes the encoding in Algorithm~\ref{algorithm:fovdetection-mpc-oblivious} unsuitable for GC is that the computation in lines~\mpclref{alg:fovmpc:nearbylat}-\mpclref{alg:fovmpc:ResultAdd} is iterated \textit{over all $n$ drones}. Because garbled circuits require a garbling table to represent every intermediate wire, the size of the circuit is large. The circuit gets larger with larger values of $n$, as the loop is unrolled to create the garbled circuit. 

Unlike our secret-shared implementation atop MOTION, however, traditional Yao-style garbled circuits work on Boolean representations. Thus, in a GC-based implementation atop EMP, there is no need to avoid Boolean comparisons in the MPC computation (\textit{cf.}~the discussion under ``Avoiding Boolean comparisons in MPC'' in \sectref{sec:pdmpc:mpc}). As a result, one can encode the computation so that the complex dot product calculations are performed only on drones that are in the vicinity of the citizen. We show the resulting encoding in Algorithm~\ref{algorithm:fovdetection-mpc-gc}. Line~\gclref{alg:fovgc:IfVicinityCheck} ensures that the computation required for field-of-view determination is only performed on drones in the vicinity of the citizen.

\begin{algorithm}[ht!]
\SetAlgoLined
\SetNlSty{}{\GCALGOLINEPREFIX}{}
\LinesNumbered
\DontPrintSemicolon
\footnotesize
\KwIn{From citizen: $\latitude_c$, $\longitude_c$, \textsf{LatVicinity}, \textsf{LongVicinity}.}
\KwIn{From regulatory authority, identity, position and $\vec{\mathbf{D}}$ vector for all drones: (\textsf{Id}$^i$, $\latitude_t^i$, $\longitude_t^i$, $\vec{\mathbf{D}}^i$) for all drones in city ($i$=$1$$\ldots$$n$).} 
\KwIn{Publicly-known: $\theta_1$, $\ldots$, $\theta_n$ for all drones.}
\KwOut{\textit{Revealed to citizen:} 
$\langle\textsf{DotP}^i, \textsf{NormSquare}^i\rangle$ for drones \textit{\underline{in vicinity}}.}
\textsf{[$\latitude$diff]} = \funcname{LatitudeDiff}($\latitude_t^1$, $\ldots$, $\latitude_t^n$, $\latitude_c$)\; \label{alg:fovgc:latdiff}
\textsf{[$\longitude$diff]} = \funcname{LongitudeDiff}($\longitude_t^1$, $\ldots$, $\longitude_t^n$, $\longitude_c$)\; \label{alg:fovgc:longdiff}
\For{$i~\gets$~$1$~\KwTo~$n$ \label{alg:fovgc:loopinmpc}}
{
  \If{\upshape (\textsf{$\latitude$diff[$i$]} $\leq$ \textsf{LatVicinity} \textbf{and} \textsf{$\longitude$diff[$i$]} $\leq$ \textsf{LongVicinity}) \label{alg:fovgc:IfVicinityCheck}}
  {
      $\vec{\mathbf{C}}^i$ = \funcname{Vectorize}($\latitude_t^i$, $\longitude_t^i$, $\latitude_c$, $\longitude_c$)\; \label{alg:fovgc:vec}
      
      $\textsf{DotP}^i$ = $\vec{\mathbf{D}^i}\cdot\vec{\mathbf{C}^i}$\; \label{alg:fovgc:dotp}
      
      $\textsf{NormSquare}^i$ =  
          ($\lvert\vec{\mathbf{D}^i}\rvert$)$^2$ $\times$ ($\lvert\vec{\mathbf{C}^i}\rvert$)$^2$\; \label{alg:fovgc:normsquare}
    
      \textsf{Result}.add($\langle\textsf{Id}^i, \textsf{DotP}^i, \textsf{NormSquare}^i\rangle$)\; \label{alg:fovgc:resultadd}
  }
}
\Return \textsf{Result} (\textbf{\textit{revealed only to citizen}})

\par\vskip.5\baselineskip\hrule height 0.2pt\par\vskip.5\baselineskip

\tcp{\textrm{\textbf{Steps below happen on citizen's phone (not as MPC)}}}

\For{\upshape \textbf{each} $\langle\textsf{Id}^i, \textsf{DotP}^i, \textsf{NormSquare}^i\rangle$ $\in$ \textsf{Result}}
{
    $\phi_i$ = $\arccos$($\textsf{DotP}^i$/$\sqrt{\textsf{NormSquare}^i}$)\;
    \lIf{\upshape ($\phi_i$ $\leq$ $\theta_i$)}{shortlist the drone \textsf{Id}$^i$.}
}
\caption{\funcname{DetectFieldOfView} adapted for MPC using Yao-style~\cite{yao:focs86} garbled circuits.}
\label{algorithm:fovdetection-mpc-gc}
\end{algorithm}

\begin{figure*}
\centering
\footnotesize
\setlength{\tabcolsep}{4pt}
\begin{tabular}{p{9mm}rrrrrrrr}
\thickhline
\rowcolor{LightGray}
\textbf{\scriptsize Number} & 
      \multicolumn{8}{c}{\scriptsize \textbf{$\leftarrow$ Number of drones in citizen's vicinity $\rightarrow$}}\\
\rowcolor{LightGray}
\textbf{\scriptsize in city  $\downarrow$}
          &    \multicolumn{2}{c}{\textbf{1}}       
          &    \multicolumn{2}{c}{\textbf{2}}        
          &    \multicolumn{2}{c}{\textbf{5}}         
          &    \multicolumn{2}{c}{\textbf{10}}\\
\hline
\textbf{100}       
          & \lan: & 81.6$\pm$19.7ms          
          & \lan: & 94.4$\pm$9.6ms                    
          & \lan: & 133.8$\pm$6.1ms                  
          & \lan: & 163.0$\pm$10.1ms\\
          & \wan: & 5.9$\pm$0.4s            
          & \wan: & 9.7$\pm$1.0s                 
          & \wan: & 15.8$\pm$1.4s                    
          & \wan: & 24.0$\pm$3.9s\\
\rowcolor{LightGray}
\textbf{200}       
          & \lan: & 107.2$\pm$4.7ms
          & \lan: & 127.2$\pm$19.2ms              
          & \lan: & 137.6$\pm$10.3ms                    
          & \lan: & 196.4$\pm$20.7ms\\
\rowcolor{LightGray}
          & \wan: & 7.2$\pm$0.5s             
          & \wan: & 8.1$\pm$1.5s              
          & \wan: & 15.7$\pm$1.8s
          & \wan: & 23.5$\pm$1.7s\\
\textbf{500}       
          & \lan: & 165.8$\pm$12.8ms
          & \lan: & 167.2$\pm$27.0ms   
          & \lan: & 221.6$\pm$29.6ms   
          & \lan: & 235.8$\pm$25.9ms\\
          & \wan: & 9.1$\pm$0.7s
          & \wan: & 12.5$\pm$2.1s
          & \wan: & 19.7$\pm$3.7s
          & \wan: & 21.4$\pm$2.1s\\
\rowcolor{LightGray}
\textbf{1000}      
          & \lan: & 370.2$\pm$19.6ms
          & \lan: & 383.4$\pm$25.8ms    
          & \lan: & 457.2$\pm$27.6ms
          & \lan: & 554.6$\pm$51.8ms\\
\rowcolor{LightGray}
          & \wan: & 11.0$\pm$1.5s
          & \wan: & 14.8$\pm$3.3s   
          & \wan: & 19.6$\pm$1.5s
          & \wan: & 51.9$\pm$6.2s\\
\thickhline
\end{tabular}\\
\begin{tabular}{p{\linewidth}}
{\small \textit{
Average RTT between the citizen and regulatory authority on our 1Gbps \lan\ is 0.220ms (measured with \texttt{ping}). In \wan, the citizen is connected via a 4G mobile data network, and the regulatory authority is an instance on Azure US West-Central. The average RTT is 282ms, and the citizen's data upload speed during this experiment varied between 74KBps to 328KBps, measured with SpeedTest~\cite{speedtest}.}}
\end{tabular}
\negspaceqtr
\caption{Overall query latency at citizen's mobile device (for Algorithm~\ref{algorithm:fovdetection-mpc-gc}).}
\label{figure:latency-citizen:GC}
\negspacehalf
\end{figure*}

We designate the citizen as the circuit garbler and the regulatory authority as the circuit evaluator. This is because the citizen can pre-compute a large number of garbled circuits and send to the regulatory authority, \eg~the citizen could do so periodically, and transmit these garbled circuits when connected with sufficient network bandwidth. The live network cost between the garbler and the evaluator is only the OT cost of transmitting the garbler's inputs to the circuit. Thus, during circuit evaluation, only the citizen's inputs need to be transmitted to the regulatory authority via OT, keeping the communication requirements modest when the citizen's mobile device is on a low-bandwidth, high-latency mobile WAN. In contrast, if the regulatory authority were designated as the circuit garbler and the citizen's mobile phone as the circuit evaluator, then the OT will involve transferring city-scale data of all the drones to the mobile device on each query.

However, by only executing the computation on drones in the vicinity of the citizen, Algorithm~\ref{algorithm:fovdetection-mpc-gc} has a subtle side-channel.
The side-channel arises as a result of the control-dependency between lines~\gclref{alg:fovgc:vec}-\gclref{alg:fovgc:resultadd} and the conditional on line~\gclref{alg:fovgc:IfVicinityCheck}. The side-channel is akin to implicit flows in the information-flow literature~\cite{denning:cacm77}. The regulatory authority can infer (using timing differences or the amount of network communication) the \textit{number} of drones in the citizen's vicinity, but not their \textit{identities}.

There are known methods in the literature to suppress this implicit information-flow side-channel ensuring that the conditional on line~\gclref{alg:fovgc:IfVicinityCheck} of Algorithm~\ref{algorithm:fovdetection-mpc-gc} executes as an \textit{oblivious if statement}, in which dummy statements are suitably introduced so that some blocks of code are executed regardless of the value of the conditional~\cite{oblivc:eprint}. While some MPC frameworks provide such an oblivious if statement construct (\eg~Obliv-C~\cite{oblivc:eprint}), many others do not (\eg~EMP does not), and the onus is on the algorithm designer to ensure that both branches of the conditional contain identical amount of computation (with identical timing and network communication). Moreover, making the conditional oblivious has the drawback of requiring dummy code blocks to be executed regardless of the conditional, and the encoding of these code blocks adds to the circuit size. This brings us back to the same problems that plague encoding Algorithm~\ref{algorithm:fovdetection-mpc-oblivious} using garbled circuits, as discussed earlier in this section.

As a result, Algorithm~\ref{algorithm:fovdetection-mpc-gc} chooses not to make the conditional on line~\gclref{alg:fovgc:IfVicinityCheck} oblivious. Unfortunately, this is the cost of having a practical GC implementation.  As discussed earlier, the costs are prohibitive otherwise in GC---using an oblivious if statement in Algorithm~\ref{algorithm:fovdetection-mpc-gc} with code similar in timing/communication characteristics to those in lines~\gclref{alg:fovgc:vec}-\gclref{alg:fovgc:resultadd} results in a garbled circuit that is about \tabdataref{685MB} for $n$=1000 drones.

The cost of having a non-oblivious if statement is that the regulatory authority learns the number of drones in the vicinity of the querying citizen via the resulting side-channel (but not the drone identities). Thus, even if the regulatory authority were to learn the number of drones that pass the if-conditional on line~\ref{alg:fovgc:IfVicinityCheck}, it would just learn the number of drones in the vicinity of \textit{some} citizen (and not a \textit{particular} citizen, because citizens submit queries anonymously). This is a weaker privacy guarantee than the one provided by Algorithm~\ref{algorithm:fovdetection-mpc-oblivious}, which runs the computation on all drones. For example, it may be possible for the regulatory authority to use live drone traffic density maps to estimate the approximate locations from where the citizen issues queries, which may be unacceptable in some settings (\eg~see Abidi \etal~\cite{abidi:ndss22} for such an attack).

We now present an evaluation of Algorithm~\ref{algorithm:fovdetection-mpc-gc} (implemented in EMP~\cite{wang:emp:2016}) using the same experimental setup described in \sectref{sec:evaluation:pdmpc}. As before, we measured the mobile data consumption on the citizen's device to pose queries to the regulatory authority. We also measured the end-to-end latency observed by the citizen when a query is issued. \figref{figure:latency-citizen:GC} and \figref{figure:mobile-data-usage:GC} report the results of our experiments (average and standard deviation reported over 5 runs). 

\begin{wrapfigure}{l}{0.58\linewidth}
\footnotesize
\begin{tabular}{p{9mm}cccc}
\thickhline
\rowcolor{LightGray}
\textbf{\scriptsize Number}      & 
      \multicolumn{4}{c}{\scriptsize \textbf{$\leftarrow$ No. of drones in citizen's vicinity $\rightarrow$}}\\
\rowcolor{LightGray}
\textbf{\scriptsize in city  $\downarrow$}
          &     ~~~~ \textbf{1} ~~~~~~       
          &    ~~~~~ \textbf{2} ~~~~~~        
          &    ~~~~~ \textbf{5} ~~~~~~  
          &    ~~~~~ \textbf{10}\\
\hline
\textbf{100}       
          & 1.349       & 2.032        & 4.076        & 7.483\\

\rowcolor{LightGray}
\textbf{200}      
         & 1.750       & 2.432        & 4.476         & 7.884\\
\textbf{500}      
         & 3.207       & 3.888        & 6.190         & 9.598\\
\rowcolor{LightGray}
\textbf{1000}      
          & 5.721       & 6.403       & 8.448   & 12.112\\
\thickhline
\end{tabular}
\negspaceqtr
\caption{Per-query mobile data usage (for Algorithm~\ref{algorithm:fovdetection-mpc-gc}).}
\label{figure:mobile-data-usage:GC}
\negspacehalf
\end{wrapfigure}

Observe that the mobile data consumed depends upon \textit{two} factors: the number of drones deployed in the city and the number of drones in the citizen's vicinity. The former determines the size of the garbled circuit that is generated and must be sent from the citizen to the regulatory authority (\ie~the value of $n$ in Algorithm~\ref{algorithm:fovdetection-mpc-gc}). The latter determines the number of drones for which lines~\gclref{alg:fovgc:vec}-\gclref{alg:fovgc:resultadd} of Algorithm~\ref{algorithm:fovdetection-mpc-gc} execute.  Thus, for these experiments we crafted inputs to our MPC implementation 
that vary the number of drones deployed in the city, and simulate a given drone density in the citizen's vicinity.

\figref{figure:mobile-data-usage:GC} shows that for a given density of drones in the citizen's vicinity, mobile data usage increases as the number of drones in the city-wide deployment ($n$) increases. This is because the size of the circuit to be garbled increases proportionally with $n$, thereby resulting in a larger amount of data to be sent from the citizen's phone to the regulatory authority. The mobile data usage also includes the OT time to send the citizens inputs, however, that is a constant value for all cases. In the worst case that we simulated---1000 drones city-wide, with a dense presence of 10 drones in the citizen's vicinity---each query from the citizen's phone consumes about 12.112MB of mobile data. 

This mobile data usage is roughly double  compared to the mobile data usage for 1000 drones as reported in \figref{figure:mobile-data-usage}. Also, note that \figref{figure:mobile-data-usage} reports the number for the \textit{oblivious} algorithm (Algorithm~\ref{algorithm:fovdetection-mpc-oblivious}) that executes the computation on \textit{all 1000 drones}. In contrast, the complex computation in the GC implementation of Algorithm~\ref{algorithm:fovdetection-mpc-gc} runs \textit{on only 10 drones} that clear the conditional on line~\gclref{alg:fovgc:IfVicinityCheck}. This experiment clearly justifies our use of secret-sharing based MPC in \pdmpc\ over a garbled circuit-based implementation.

For a given city-scale drone deployment (\eg~$n$=100 drones), observe that the per-query mobile data usage increases in proportion to the drone density in the citizen's vicinity. The reader may question why we observe this trend, given that the size of the garbled circuit is fixed by the value of $n$, and does not depend on drone density. The answer to this question lines in how Algorithm~\ref{algorithm:fovdetection-mpc-gc} is implemented atop EMP, as discussed below.

Recall that EMP lacks default support for oblivious if statements, and relies on the programmer to suitably design the algorithm to ensure that any side channels due to the lack of this support are suppressed. In particular, EMP will not compile Algorithm~\ref{algorithm:fovdetection-mpc-gc} (as shown) because the predicate used in the conditional statement on line~\gclref{alg:fovgc:IfVicinityCheck} is input-dependent. EMP compiles the circuit only if the value of the predicate on line~\gclref{alg:fovgc:IfVicinityCheck} is revealed to the citizen (thereby revealing to the citizen the identities of drones in their vicinity). Further, EMP transmits the garbled tables for the wires of the circuit corresponding to lines~\gclref{alg:fovgc:vec}-\gclref{alg:fovgc:resultadd} of Algorithm~\ref{algorithm:fovdetection-mpc-gc} on-the-fly, \ie~as and when the circuit evaluator exercises that branch of the computation, and not in advance.

Moreover, EMP does not directly encode the for-loop on line~\gclref{alg:fovgc:loopinmpc} in the circuit, but chooses to create a circuit with the loop unrolled (which it can, because the value of $n$ is a constant that is known to the circuit garbler). Thus, a different set of garbled tables is transmitted each time the computation executes lines~\gclref{alg:fovgc:vec}-\gclref{alg:fovgc:resultadd}. As a result, the per-query network communication increases with the drone density. This is indeed a side-channel that allows the regulatory authority to determine the drone density in the anonymously querying citizen's vicinity (but not their identities). In summary, the various performance optimizations that are required to make a GC-based realization of Algorithm~\ref{algorithm:fovdetection-mpc-gc} practical in terms of network communication overhead in turn result in an implementation that offers weaker privacy guarantees than the secret-shared implementation of Algorithm~\ref{algorithm:fovdetection-mpc-oblivious}, implemented in our \pdmpc\ prototype.

\section{Security Analysis of \pdros}
\label{sec:appendix:pdros-analysis}

This section analyzes the security of the \pdros\ design described in \sectref{sec:pdros:main}. Recall that our design is tailored to allow well-intentioned ROS applications to establish that they comply with privacy by: \bfcircled{a}~ensuring that sensitive data that they publish are consumed only by the trusted sanitizer application running in the secure world; and \bfcircled{b}~ensuring that they only subscribe to data that has been sanitized by the trusted video sanitizer.

\pdros\ relies on two key security mechanisms:
\begin{mylist}
\item \textit{Attestation reports.} \pdros\ relies on boot-time attestation as well as runtime attestation of applications as they are launched. Boot-time attestation ensures the integrity of the secure world when the drone is initially booted up. The secure world also performs an integrity measurement of the normal world when it boots the normal world. Further, the design of \pdros\ measures the integrity of the normal world, ROS, SROS, Sanitizer-FE as well as the application launcher and stores the corresponding measurement in the secure world via the trusted audit logger. A security analyst can use these measurements in the audit trail to determine that the normal world, ROS, SROS, Sanitizer-FE and the application launcher are untampered at least as of the time of measurement, \ie~when an application is launched in the normal world. We rely, as is standard in all attestation protocols~\cite{tpm:sec2004}, on the integrity measurements being rooted (transitively) in trusted hardware, in our case, the ARM TrustZone.
\item \textit{Verifying flow redirection.} The application launcher stores a copy of the manifest of the application that it launches in the secure world. SROS enforces that all ROS applications communicate in accordance with their declared manifests. Manifests are part of each application's X.509 certificates, and the SROS application launcher verifies the certificate of the application during launch, thereby ensuring that neither the application nor the manifest have been tampered with. 
\end{mylist}

Once the manifests are stored in the audit trail, a security analyst can use the stored manifests to determine that: 
\bfcircled{a}~if an application publishes sensitive data, which we identify by their topic names, \eg~\rostopic{VideoFeed}, then no application other than the trusted sanitizer's front-end subscribes to that topic; and
\bfcircled{b}~if an application must consume sensitive data, then it only consumes the sanitized data stream published by the trusted video sanitizer, \eg~\rostopic{PrivVideoFeed}. 

Given the above discussion, the security provided by \pdros\ relies on the following factors:
\begin{mybullet}
\item (\goal{F1}) \textit{Reliability of attestation reports.} \pdros\ verifies the integrity of the normal world, ROS, SROS, Sanitizer-FE and the application launcher via the attestation reports. However, trusted hardware-based attestation protocols are known to be vulnerable to time of check to time of use (TOCTTOU) problems, and our \pdros\ prototype is no different. An attacker can use zero-day exploits against either the normal world, ROS, SROS, Sanitizer-FE or the application launcher causing their runtime behavior to differ from that of the attested version. As a result, it may be possible to bypass SROS enforcement of application manifests using such TOCTTOU attacks. That said, \pdros\ can incorporate recently-proposed methods to address the TOCTTOU problem in remote attestation~\cite{rata:ccs2021}. \pdros\ could also use fine-grained path-based attestation of application functionality (\eg\ C-FLAT~\cite{cflat:ccs2016} or OAT~\cite{oat:oak2020}), which attests the precise execution path followed by the application, thereby providing protection against runtime exploits on the application.
\item (\goal{F2}) \textit{Faithfulness of manifests in capturing inter-application communication.} \pdros\ relies in a key way on the runtime enforcement of application communication patterns based on the publish/subscribe topics declared in the SROS application manifests. This assumption is satisfied so long as ROS applications only communicate via ROS abstractions, \ie~topics. However, it is known~\cite{privaros:ccs2020} that applications can bypass ROS abstractions, and communicate directly via low-level abstractions, \eg~sockets or shared memory, by invoking raw system calls. The ROS community has addressed this problem by building MAC enforcement of the application communication patterns implied by the manifests within the operating system at the level of processes implementing the ROS applications (\eg~the Privaros~\cite{privaros:ccs2020} or the SROS+AppArmor\cite{ros+apparmor} projects). \pdros\ could also use these methods to enhance the normal world operating system, and obtain protection from attacks that bypass ROS communication abstractions. 

An alternative approach is for an auditing authority to verify the ROS applications running on the drone via static analysis. The drone operator can provide the application binaries to the citizen (or any other auditing authority), who then verifies, using the attestation reports, that those were the binaries that were launched on the drone. The auditing authority can then use static analysis on each application binary to ensure the absence of low-level calls (\eg~system calls to open sockets or establish shared memory) in the application binary. Applications that pass this analysis can therefore only use SROS for communication, and the corresponding SROS manifests therefore suffice to capture inter-application communication patterns. This approach has the advantage of imposing no additional runtime performance overheads, unlike the MAC enforcement systems discussed in the previous paragraph.
\item (\goal{F3}) \textit{Ability of trusted sanitizers to enforce privacy.} Finally, \pdros\ relies on domain-specific sanitizers to enforce privacy. For example, it relies on the video sanitizer to locate all privacy-sensitive objects in each frame (\eg~human faces or vehicle registration plates), and suitably distort them so that the sanitized feed respects privacy. Since the trusted sanitizer is part of the trusted-computing base, we do not attempt to further verify the functioning of the sanitizer in our \pdros\ design. A pair of malicious ROS application could also attempt to leak privacy-sensitive objects  via a low-rate side-channel that still passes through the trusted sanitizer. For example, a malicious source application could signal the presence of a particular individual in the video feed using a single bit in the header of the video stream. If this bit is not detected and sanitized by the trusted sanitizer, a colluding malicious downstream target application can infer the presence of the individual as indicated by the malicious source application. Such side-channel attacks are currently out of scope for \pdros, and new methods need to be developed to detect such side-channels.
\end{mybullet}

Finally, we note that the security analysis above applies to the design of \pdros\ discussed in \sectref{sec:pdros:main}, which was presented assuming that the underlying ARM TrustZone SoC lacks support for the TrustZone Peripheral Controller (TZPC)~\cite{arm:tzpc:nov04}. If the SoC indeed supports TZPC, then \pdros\ can use the design outlined in \sectref{sec:pdros:goals}, with sensitive data confined to the secure world, and only leaving the secure world in sanitized form. With this TZPC-based design, the factors \goal{F1} and \goal{F2} discussed above do not apply (disregarding zero-day exploits against the secure world itself, as is standard, because it is part of the trusted-computing base). In this design, the only security consideration would be factor \goal{F3}, \ie~that of ensuring that the trusted sanitizer indeed removes all occurrences of sensitive objects from the video feed. It must not come as a surprise that the \pdros\ design that leverages TZPC hardware support offers stronger security guarantees (or relies on the security of fewer components) than when such support is not available.





\begin{thebibliography}{10}

\bibitem{3deo}
{3DEO}.
\newblock Rogue drone detection and mitigation.
\newblock \url{https://3deo.biz/applications/drone-detection-and-mitigation}.

\bibitem{cflat:ccs2016}
T.~Abera, N.~Asokan, L.~Davi, J.~Ekberg, T.~Nyman, A.~Paverd, A-R. Sadeghi, and
  G.~Tsudik.
\newblock {C-FLAT: Control-Flow} attestation for embedded systems software.
\newblock In {\em ACM Conference on Computer and Communications Security},
  2016.

\bibitem{abidi:ndss22}
I.~Abidi, I.~Nangia, P.~Aditya, and R.~Sen.
\newblock {Privacy in Urban Sensing with Instrumented Fleets, Using Air
  Pollution Monitoring As A Usecase}.
\newblock In {\em Network and Distributed Systems Security Symposium}, 2022.

\bibitem{ipic:mobisys2016}
P.~Aditya, R.~Sen, P.~Druschel, S-J. Oh, R.~Benenson, M.~Fritz, B.~Schiele,
  B.~Bhattacharjee, and T.~T. Wu.
\newblock {I-Pic: A Platform for Privacy-Compliant Image Capture}.
\newblock In {\em ACM Conference on Mobile Systems, Applications, and
  Services}, 2016.

\bibitem{amazon:prime:firstdelivery}
Amazon~Prime Air, December 2016.
\newblock
  \url{https://www.amazon.com/Amazon-Prime-Air/b?ie=UTF8&node=8037720011}.

\bibitem{aly:eprint:2019}
A.~Aly and N.~P. Smart.
\newblock Benchmarking privacy preserving scientific operations.
\newblock In {\em Cryptology ePrint Archive: Report 2019/354}, April 2019.

\bibitem{ros+apparmor}
{AppArmor and ROS}.
\newblock \url{http://wiki.ros.org/SROS/Tutorials/AppArmorAndROS}.

\bibitem{arm:tzpc:nov04}
ARM.
\newblock {PrimeCell$^\textsuperscript{\textregistered}$ Infrastructure
  AMBA$^\textsuperscript{TM}$ 3 TrustZone Protection Controller
  (BP147)---Revision: r0p0---Technical Overview}, November 2004.

\bibitem{arm2009security}
ARM.
\newblock Security technology building a secure system using {TrustZone}
  technology (white paper).
\newblock {\em ARM Limited}, 2009.
\newblock
  \url{https://community.arm.com/cfs-file/__key/telligent-evolution-components-attachments/01-2057-00-00-00-00-53-99/PRD29_2D00_GENC_2D00_009492C_5F00_trustzone_5F00_security_5F00_whitepaper.pdf}.

\bibitem{uk:caa}
United Kingdom-Civil~Aviation Authority.
\newblock {The Drone and Model Aircraft Code---Protecting people’s privacy}.
\newblock
  \url{https://register-drones.caa.co.uk/drone-code/protecting-peoples-privacy}.

\bibitem{faa:remote-id:2021}
United States Federal~Aviation Authority.
\newblock {{RemoteID}: Remote Identification of Unmanned Aircraft}, 15th
  January 2021.
\newblock
  \url{https://www.federalregister.gov/documents/2021/01/15/2020-28948/remote-identification-of-unmanned-aircraft}.

\bibitem{knox:ccs14}
A.~Azab, P.~Ning, J.~Shah, Q.~Chen, R.~Bhutkar, G.~Ganesh, J.~Ma, and W.~Shen.
\newblock Hypervision across worlds: {Real-time} kernel protection from the
  {ARM TrustZone} secure world.
\newblock In {\em ACM Conference on Computer and Communications Security},
  2014.

\bibitem{airbus:blueprint:2018}
K.~Balakrishnan, J.~Polastre, J.~Mooberry, R.~Golding, and P.~Sachs.
\newblock {Blueprint for the Sky---{The} roadmap for the safe integration of
  autonomous aircraft}, 2018.
\newblock \url{https://www.airbusutm.com/uam-resources-airbus-blueprint}.

\bibitem{privaros:ccs2020}
R.~Beck, A.~Vijeev, and V.~Ganapathy.
\newblock {Privaros: {A} Framework for Privacy-Compliant Delivery Drones}.
\newblock In {\em ACM Conference on Computer and Communications Security},
  2020.

\bibitem{nassi:sp21}
R.~Ben-Netanel, B.~Nassi, A.~Shamir, and Y.~Elovici.
\newblock {Detecting Spying Drones}.
\newblock {\em IEEE Security and Privacy}, 19(1), 2021.

\bibitem{birnbach:ndss:17}
S.~Birnbach, R.~Baker, and I.~Martinovic.
\newblock Wi-fly?: Detecting privacy invasion attacks by consumer drones.
\newblock In {\em Network and Distributed Systems Security Symposium}, 2017.

\bibitem{japan:apr15}
D.~Bolton.
\newblock {Man arrested for landing `radioactive' drone on Japanese Prime
  Minister's roof}.
\newblock In {\em The Independent}, April 24 2015.

\bibitem{motion:tops22}
L.~Braun, D.~Demmler, T.~Schneider, and O.~Tkachenko.
\newblock {MOTION: A Framework for Mixed-Protocol Multi-Party Computation}.
\newblock {\em ACM Transactions on Privacy and Security}, 8:8:1--8:35, May
  2022.

\bibitem{busset15}
J.~Busset, F.~Perrodin, P.~Wellig, B.~Ott, K.~Heutschi, T.~Rühl, and
  T.~Nussbaumer.
\newblock Detection and tracking of drones using advanced acoustic cameras.
\newblock {\em Unmanned/Unattended Sensors and Sensor Networks XI and Advanced
  Free-Space Optical Communication Techniques and Applications}, 2015.

\bibitem{case:naecon08}
E.~E. Case, A.~M. Zelnio, and B.~D. Rigling.
\newblock Low-cost acoustic array for small {UAV} detection and tracking.
\newblock In {\em IEEE National Aerospace \& Electronics Conference}, 2008.

\bibitem{spiders:chi2017}
V.~Chang, P.~Chundury, and M.~Chetty.
\newblock ``{Spiders} in the sky'': {User} perceptions of drones, privacy, and
  security.
\newblock In {\em ACM SIGCHI Conference on Human Factors in Computing Systems},
  2017.

\bibitem{wsj:wing-coffee:2019}
Mike Cherney.
\newblock {Some Want Delivery Drones to Buzz Off. {Would} Stricter Rules Change
  Their Minds?}
\newblock {\em Wall Street Journal}, 2nd August 2019.

\bibitem{church:optics18}
P.~Church, C.~Grebe, J.~Matheson, and B.~Owens.
\newblock Aerial and surface security applications using {LIDAR}.
\newblock In {\em Laser Radar Technology and Applications XXIII---International
  Society for Optics and Photonics}, volume 10636, 2018.

\bibitem{dsp:amazon:cbnc}
CNBC.
\newblock Amazon says this business opportunity could make you up to \$300k a
  year---here's how to get into the program, September 2018.
\newblock
  \url{https://www.cnbc.com/2018/09/06/amazon-delivery-service-partner-program-gets-thousands-of-applications.html}.

\bibitem{silver:crypto21}
G.~Couteau, P.~Rindal, and S.~Raghuraman.
\newblock {Silver: Silent VOLE and Oblivious Transfer from Hardness of Decoding
  Structured LDPC Codes}.
\newblock In {\em CRYPTO -- Advances in Cryptology}, 2021.

\bibitem{covington:zebra:2020}
Taylor Covington.
\newblock {Could Delivery Drones Be the Next Tech Privacy Violation? 88\% of
  Americans Think So}.
\newblock {\em The Zebra}, April 2020.
\newblock
  \url{https://www.thezebra.com/resources/home/delivery-drones-survey/}.

\bibitem{iraq:nov21}
J.~Davison and A.~Rasheed.
\newblock {Iraqi PM safe after drone attack on residence, military says}.
\newblock In {\em Reuters}, November 7 2021.

\bibitem{DDS_doc}
{Data Distribution Service (DDS)}.
\newblock \url{https://www.omg.org/spec/DDS/1.4/PDF}.

\bibitem{swiss:dronerules:2021}
Weibe de~Jager.
\newblock Switzerland launches world’s first {Remote ID} network for drones,
  September 2021.
\newblock
  \url{https://www.dronewatch.eu/switzerland-launches-worlds-first-remote-id-network-for-drones/}.

\bibitem{aby:ndss15}
D.~Demmler, T.~Schneider, and M.~Zohner.
\newblock {ABY--A Framework for Efficient Mixed-Protocol Secure Two-Party
  Computation}.
\newblock In {\em Network and Distributed Systems Security Symposium}, 2015.

\bibitem{denning:cacm77}
D.~E. Denning and P.~J. Denning.
\newblock Certification of programs for secure information flow.
\newblock {\em Communications of the ACM}, 20(7), July 1977.

\bibitem{npr:dec2020}
J.~Diaz.
\newblock {Amazon, TikTok, Facebook}, others ordered to explain what they do
  with user data.
\newblock {\em National Public Radio}, December 2020.
\newblock
  \url{https://www.npr.org/2020/12/15/946583479/amazon-tiktok-facebook-others-ordered-to-explain-what-they-do-with-user-data}.

\bibitem{dieber2017security}
B.~Dieber, B.~Breiling, S.~Taurer, S.~Kacianka, S.~Rass, and P.~Schartner.
\newblock {Security for the Robot Operating System}.
\newblock {\em Robotics and Autonomous Systems}, 98, 2017.

\bibitem{dieber2016application}
B.~Dieber, S.~Kacianka, S.~Rass, and P.~Schartner.
\newblock {Application-level security for ROS-based applications}.
\newblock In {\em Intelligent Robots and Systems (IROS), 2016 IEEE/RSJ
  International Conference on}. IEEE, 2016.

\bibitem{drones4sec}
{Drones4Sec -- European Federation -- General Secretariat by ZENON7}, June
  2021.
\newblock \url{https://www.drones4sec.eu/}.

\bibitem{dsp:flipkart}
{How To Get Flipkart Franchise}.
\newblock \url{https://www.steptowardbusiness.com/flipkart-franchise}.

\bibitem{cnn:street-view}
Clare Duffy.
\newblock {Google agrees to pay \$13 million in {Street View} privacy case},
  2019.
\newblock CNN Business, July 25, 2019,
  \url{https://edition.cnn.com/2019/07/22/tech/google-street-view-privacy-lawsuit-settlement/index.html}.

\bibitem{eshel13}
T.~Eshel.
\newblock Mobile radar optimized to detect {UAV}s, precision guided weapons.
\newblock {\em Defense Update}, 2013.

\bibitem{swiggy:et:2022}
ETtech.
\newblock {Swiggy} to pilot drone-based deliveries for its grocery service
  {Instamart}.
\newblock {\em Economic Times}, April 2022.
\newblock
  \url{https://m.economictimes.com/tech/startups/swiggy-to-pilot-drone-based-deliveries-for-its-grocery-service-instamart/amp_articleshow/91188144.cms}.

\bibitem{eu:dronerules:2019}
Commission Implementing~Regulation (EU).
\newblock Rules and procedures for the operation of unmanned aircraft, May
  2019.
\newblock
  \url{https://ec.europa.eu/info/law/better-regulation/have-your-say/initiatives/1642-Detailed-rules-on-unmanned-aircrafts_en}.

\bibitem{garg:hotmobile2020}
N.~Garg and N.~Roy.
\newblock Acoustic sensing for detecting projectile attacks on small drones.
\newblock In {\em HotMobile'20: 21st International Workshop on Mobile Computing
  Systems and Applications}, 2020.

\bibitem{sprobes:2014}
X.~Ge, H.~Vijayakumar, and T.~Jaeger.
\newblock {\textsc{Sprobes}: Enforcing Kernel Code Integrity on the TrustZone}.
\newblock In {\em IEEE Workshop on Mobile Security Technologies}, 2014.

\bibitem{gmw:stoc87}
O.~Goldreich, S.~Micali, and A.~Wigderson.
\newblock {How to Play any Mental Game, or A Completeness Theorem for Protocols
  with Honest Majority}.
\newblock In {\em ACM Symposium on the Theory of Computing}, 1987.

\bibitem{trusty}
Google.
\newblock {Trusty TEE -- Android Open Source Project}.
\newblock \url{https://source.android.com/security/trusty}.

\bibitem{digsky}
{Government of India}.
\newblock {Digital Sky Portal --- Office of the Director General of Civil
  Aviation}.
\newblock \url{https://digitalsky.dgca.gov.in/}.

\bibitem{digskyR1E1}
{Government of India}.
\newblock {Office of the Director General of Civil Aviation: DGCA RPAS Guidance
  Manual, Revision One of First Edition}, June 2019.
\newblock
  \url{https://diceindia.org.in/wp-content/uploads/Updated-DGCA-RPAS-Guidance-Manual.pdf}.

\bibitem{gpstodistance}
{Approximate Metric Equivalents for Degrees, Minutes, and Seconds}, August
  2019.
\newblock
  \url{https://www.usna.edu/Users/oceano/pguth/md_help/html/approx_equivalents.htm}.

\bibitem{dds}
Object~Management Group.
\newblock {About the Data Distribution Service Specification Version 1.4}.
\newblock \url{https://www.omg.org/spec/DDS/About-DDS/}.

\bibitem{koettl:nytimes:2018}
C.~Koettl and B.~Marcolini.
\newblock A closer look at the drone attack on {Maduro in Venezuela}, August
  2018.
\newblock
  \url{https://www.nytimes.com/2018/08/10/world/americas/venezuela-video-analysis.html}.

\bibitem{l4t}
{NVIDIA Jetson Linux Driver Package (L4T)}.
\newblock \url{https://developer.nvidia.com/embedded/linux-tegra}.

\bibitem{secloak:mobisys18}
M.~Lentz, R.~Sen, P.~Druschel, and B.~Bhattacharjee.
\newblock {SeCloak: ARM Trustzone-based Mobile Peripheral Control}.
\newblock In {\em ACM Symposium on Mobile Systems, Applications, and Services},
  2018.

\bibitem{liu2017protc}
R.~Liu and M.~Srivastava.
\newblock {PROTC}: Protecting drone's peripherals through {ARM TrustZone}.
\newblock In {\em 3rd Workshop on Micro Aerial Vehicle Networks, Systems, and
  Applications}, 2017.

\bibitem{alidrone2018icdcs}
T.~Liu, A.~Hojjati, A.~Bates, and K.~Nahrstedt.
\newblock {AliDrone: Enabling Trustworthy Proof-of-Alibi for Commercial Drone
  Compliance}.
\newblock In {\em IEEE 38th International Conference on Distributed Computing
  Systems}, 2018.

\bibitem{dsp:amazon}
Amazon Logistics.
\newblock {Amazon Logistics---Delivery Services Partners Program}.
\newblock \url{https://logistics.amazon.com/marketing/opportunity}.

\bibitem{mcclean2013preliminary}
J.~McClean, C.~Stull, C.~Farrar, and D.~Mascare{\~n}as.
\newblock {A Preliminary Cyber-Physical Security Assessment of the Robot
  Operating System (ROS)}.
\newblock In {\em Unmanned Systems Technology XV}, volume 8741. International
  Society for Optics and Photonics, 2013.

\bibitem{mccollough:1893}
Ernest McCollough.
\newblock {\em {Photographic Topography}}.
\newblock Industrial Publishing Company, San Francisco, 1893.

\bibitem{nassi:oakland:19}
B.~Nassi, R.~Ben-Netanel, A.~Shamir, and Y.~Elovici.
\newblock {Drones' Cryptanalysis -- Smashing Cryptography with a Flicker}.
\newblock In {\em IEEE Symposium on Security and Privacy}, 2019.

\bibitem{nassi:cscml21}
B.~Nassi, R.~Ben-Netanel, A.~Shamir, and Y.~Elovici.
\newblock {Game of Drones - Detecting Spying Drones Using Time Domain
  Analysis}.
\newblock In {\em 5th International Symposium on Cyber-Security, Cryptology and
  Machine Learning}, 2021.

\bibitem{nassi:oakland21}
B.~Nassi, R.~Bitton, R.~Masuoka, A.~Shabtai, and Y.~Elovici.
\newblock {SoK: Security and Privacy in the Age of Commercial Drones}.
\newblock In {\em IEEE Symposium on Security and Privacy}, 2021.

\bibitem{matthan:mobisys:2017}
P.~Nguyen, H.~Truong, M.~Ravindranathan, A.~Nguyen, R.~Han, and T.~Vu.
\newblock Matthan: {Drone} presence detection by identifying physical
  signatures in the drone’s {RF} communication.
\newblock In {\em 15th Annual International Conference on Mobile Systems,
  Applications, and Services}, 2017.

\bibitem{rata:ccs2021}
I.~Nunes, S.~Jakkamsetti, N.~Rattanavipanon, and G.~Tsudik.
\newblock On the {TOCTOU} problem in remote attestation.
\newblock In {\em ACM Conference on Computer and Communications Security},
  2021.

\bibitem{australia:oaic}
Office of~the Australian Information~Commissioner.
\newblock Survellience and monitoring--drones.
\newblock
  \url{https://www.oaic.gov.au/privacy/your-privacy-rights/surveillance-and-monitoring/drones/}.

\bibitem{wing-trial:aus-parliament-report:2019}
Standing~Committee on~Economic~Development and Tourism.
\newblock {Inquiry into Drone Delivery Systems in the Australian Capital
  Territory, Report~6}, July 2019.

\bibitem{parrot:anafi-ai}
Parrot.
\newblock {ANAFI Ai---The 4G robotic UAV}, November 2021.
\newblock
  \url{https://www.parrot.com/assets/s3fs-public/2021-11/white-paper-anafi-ai-v1.6.pdf}.

\bibitem{aby2:usenix21}
A.~Patra, T.~Schneider, A.~Suresh, and H.~Yalame.
\newblock {ABY2.0: Improved Mixed-Protocol Secure Two-Party Computation}.
\newblock In {\em USENIX Security Symposium}, 2021.

\bibitem{pedinti:ubicomp11}
S.~T. Peddinti and N.~Saxena.
\newblock On the limitations of query obfuscation techniques for location
  privacy.
\newblock In {\em ACM International Conference on Ubiquitous Computing}, 2011.

\bibitem{puttaswamy:isc:2012}
{In the Supreme Court of India, Civil Original Jurisdiction, Writ Petition
  (Civil) No. 494 of 2012: Justice K. S. Puttaswamy and ANR vs. Union of India
  and ORS}, 2012.
\newblock
  \url{https://main.sci.gov.in/supremecourt/2012/35071/35071_2012_Judgement_24-Aug-2017.pdf}.

\bibitem{sounduav:dronet19}
S.~Ramesh, T.~Pathier, and J.~Han.
\newblock {SoundUAV: Towards Delivery Drone Authentication via Acoustic Noise
  Fingerprinting}.
\newblock In {\em 5th Workshop on Micro Aerial Vehicle Networks, Systems, and
  Applications}, 2019.

\bibitem{rodriguez2018message}
F.~J. Rodr{\'\i}guez-Lera, V.~Matell{\'a}n-Olivera, J.~Balsa-Comer{\'o}n,
  {\'A}.~M. Guerrero-Higueras, and C.~Fern{\'a}ndez-Llamas.
\newblock {Message Encryption in Robot Operating System: Collateral Effects of
  Hardening Mobile Robots}.
\newblock {\em Frontiers in ICT}, 5, 2018.

\bibitem{ros2}
{ROS 2}--{ROS 2} documentation, the latest version of the robot operating
  system.
\newblock \url{https://index.ros.org/doc/ros2/}.

\bibitem{rozantsev:cvpr:2015}
A.~Rozantsev, V.~Lepetit, and P.~Fua.
\newblock Flying objects detection from a single moving camera.
\newblock In {\em IEEE Conference on Computer Vision and Pattern Recognition},
  2015.

\bibitem{tpm:sec2004}
R.~Sailer, X.~Zhang, T.~Jaeger, and L.~van Doorn.
\newblock Design and implementation of a {TCG}-based integrity measurement
  architecture.
\newblock In {\em USENIX Security}, 2004.

\bibitem{911security}
911 security.
\newblock {Drone Detection by 911 Security---Detect and Track drones in your
  Airspace with the AirGuard Software Platform}.
\newblock \url{https://www.911security.com/}.

\bibitem{sinnott:1984}
R.~W. Sinnott.
\newblock {Virtues of the {Haversine}}.
\newblock {\em Sky and Telescope}, 68(2), August 1984.

\bibitem{snyder:1993}
John~P. Snyder.
\newblock {\em {Flattening the Earth: Two Thousand Years of Map Projections}}.
\newblock University of Chicago Press, 1993.

\bibitem{speedtest}
{Ookla SpeedTest}.
\newblock \url{https://www.speedtest.net}.

\bibitem{meta:reuters:2022}
Jonathan Stempel.
\newblock {Meta's Facebook} to pay \$90 million to settle privacy lawsuit over
  user tracking.
\newblock {\em Reuters News}, February 2022.
\newblock
  \url{https://www.reuters.com/technology/metas-facebook-pay-90-million-settle-privacy-lawsuit-over-user-tracking-2022-02-15/}.

\bibitem{oat:oak2020}
Z.~Sun, B.~Feng, L.~Lu, and S.~Jha.
\newblock {OAT: Attesting} operation integrity of embedded devices.
\newblock In {\em IEEE Symposium on Security and Privacy}, 2020.

\bibitem{qsystems}
Quanergy Systems.
\newblock Quanergy systems to showcase powerful lidar security detection system
  at isc west, April 2016.
\newblock
  \url{https://quanergy.com/wp-content/uploads/2020/02/Quanergy-Systems-to-Showcase-Powerful-LiDAR-Security-Detection-System-at-ISC-West-4-6-2016.pdf}.

\bibitem{thales:scaleflyt}
Thales.
\newblock {ScaleFlyt Remote ID: Identification and tracking for safe drone
  operations}.
\newblock
  \url{https://www.thalesgroup.com/en/markets/aerospace/drone-solutions/scaleflyt-remote-id-identification-tracking-safe-drone-operations}.

\bibitem{vincenty:1975}
T.~Vincenty.
\newblock {Direct and Inverse Solutions of Geodesics on the Ellipsoid with
  Application of Nested Equations}.
\newblock {\em Survey Review}, 23, 1975.

\bibitem{wang:emp:2016}
X.~Wang, A.~J. Malozemoff, and J.~Katz.
\newblock {EMP-toolkit: Efficient MultiParty computation toolkit}, 2016.
\newblock \url{https://github.com/emp-toolkit}.

\bibitem{perceptions:popets2016}
Y.~Wang, H.~Xia, Y.~Yao, and Y.~Huang.
\newblock Flying eyes and hidden controllers: {A} qualitative study of people's
  privacy perceptions of civilian drones in the {US}.
\newblock {\em Proceedings on Privacy Enhancing Technologies (PoPETS)}, 3,
  2016.

\bibitem{white2019sros1}
R.~White, G.~Caiazza, H.~Christensen, and A.~Cortesi.
\newblock {SROS1: Using and developing secure ROS1 systems}.
\newblock In {\em Robot Operating System (ROS)}. Springer, 2019.

\bibitem{white2016sros}
R.~White, D.~Christensen, I.~Henrik, and D.~Quigley.
\newblock {SROS: Securing ROS over the Wire, in the Graph, and through the
  Kernel}.
\newblock {\em arXiv:1611.07060}, 2016.

\bibitem{white2018procedurally}
R.~White, H.~Christensen, G.~Caiazza, and A.~Cortesi.
\newblock Procedurally provisioned access control for robotic systems.
\newblock In {\em IEEE/RSJ International Conference on Intelligent Robots and
  Systems}, 2018.

\bibitem{ffna:remote-id:2018}
Sarah Whittaker.
\newblock {New Draft Drone Laws for {France} Require {Remote ID} And {Signal}},
  April 2018.
\newblock
  \url{https://dronebelow.com/2018/04/12/new-draft-drone-laws-for-france-require-remote-id-and-signal/}.

\bibitem{wiki:street-view}
Wikipedia.
\newblock {Google Street View privacy concerns}.
\newblock
  \url{https://en.wikipedia.org/wiki/Google_Street_View_privacy_concerns}
  (note: contains pointers to cases in numerous countries banning or strictly
  regulating Google Street View).

\bibitem{yao:focs86}
A.~Yao.
\newblock {How to generate and exchange secrets}.
\newblock In {\em IEEE Symposium on the Foundations of Computer Science}, 1986.

\bibitem{bystanders:chi2017}
Y.~Yao, H.~Xia, Y.~Huang, and Y.~Wang.
\newblock Privacy mechanisms for drones: {Perceptions} of drone controllers and
  bystanders.
\newblock In {\em ACM SIGCHI Conference on Human Factors in Computing Systems},
  2017.

\bibitem{oblivc:eprint}
S.~Zahur and D.~Evans.
\newblock {Obliv-C: A Language for Extensible Data-Oblivious Computation}.
\newblock In {\em Cryptology ePrint Archive: Report 2015/1153}, November 2015.

\end{thebibliography}
\end{document}